\documentclass[fleqn,usenatbib]{mnras}

\usepackage{newtxtext,newtxmath}

\usepackage[T1]{fontenc}
\usepackage{ae,aecompl}
\usepackage[utf8]{inputenc}

\usepackage{graphicx}	
\usepackage{bm,tensor,braket}
\usepackage{xcolor}
\usepackage{enumerate}
\usepackage{mathtools}
\usepackage[normalem]{ulem}

\title[Multiband LIGO/Virgo BBHs]{Multiband Observation of LIGO/Virgo
Binary Black Hole Mergers in the Gravitational-wave Transient Catalog
GWTC-1}

\author[C. Liu et al.]{
Chang Liu,$^{1,2}$
Lijing Shao,$^{2}$\thanks{Corresponding author. E-mail: lshao@pku.edu.cn (LS)}
Junjie Zhao,$^{3}$
and Yong Gao$^{1,2}$
\\
$^{1}$Department of Astronomy, School of Physics, Peking University,
Beijing 100871, China\\
$^{2}$Kavli Institute for Astronomy and Astrophysics, Peking University,
Beijing 100871, China\\
$^{3}$School of Physics and State Key Laboratory of Nuclear Physics and
Technology, Peking University, Beijing 100871, China
}

\date{Accepted XXX. Received YYY; in original form ZZZ}

\pubyear{2020}

\begin{document}
\label{firstpage}
\pagerange{\pageref{firstpage}--\pageref{lastpage}}
\maketitle

\begin{abstract}
	The Advanced LIGO and Virgo detectors opened a new era to study black
	holes (BHs) in our Universe. A population of stellar-mass binary BHs
	(BBHs) are discovered to be heavier than previously expected. These
	heavy BBHs provide us an opportunity to achieve multiband observation
	with ground-based and space-based gravitational-wave (GW) detectors. In
	this work, we use BBHs discovered by the LIGO/Virgo Collaboration as
	indubitable examples, and study in great detail the prospects for
	multiband observation with GW detectors in the near future. We apply
	the Fisher matrix to spinning, non-precessing inspiral-merger-ringdown
	waveforms, while taking the motion of space-based GW detectors fully
	into account. Our analysis shows that, detectors with decihertz
	sensitivity are expected to log stellar-mass BBH signals with very
	large signal-to-noise ratio and provide accurate parameter estimation,
	including the sky location and time to coalescence. Furthermore, the
	combination of multiple detectors will achieve unprecedented
	measurement of BBH properties. As an explicit example, we present the
	multiband sensitivity to the generic dipole radiation for BHs, which is
	vastly important for the equivalence principle in the foundation of
	gravitation, in particular for those theories that predict
	curvature-induced scalarization of BHs.
\end{abstract}

\begin{keywords}
	binaries: general --  gravitational waves -- stars: black holes 
\end{keywords}



\section{Introduction}
\label{sec:intro}


The first direct detection of gravitational wave (GW) from a binary black
hole (BBH) merger by the LIGO/Virgo Collaboration \citep{Abbott:2016blz}
not only opened a new era to observe the {\it dark side} of our Universe,
but also surprised astrophysicists with black holes (BHs) whose masses are
much heavier than what were previously expected
\citep{TheLIGOScientific:2016htt}. While it calls the population synthesis
of BBHs into reconsideration and rectification, in the meantime, it
provides an opportunity for multiband GW observation \citep{Sesana:2016ljz,
Cutler:2019krq}.

In the 2030s, the Laser Interferometer Space Antenna (LISA) will begin its
GW observation \citep{Audley:2017drz}. \citet{Sesana:2016ljz} pointed out
that, heavy BBHs like the GW150914 could be observed in the millihertz band
of LISA, before they enter the hectohertz band of the Advanced LIGO/Virgo
detectors. The scientific targets of LISA to fundamental physics including
stellar-mass BBHs are summarized in \citet{Barausse:2020rsu}, with
multiband observation as an important goal. Actually, multiband detection
could benefit space and ground observation in a mutual way. Not only LISA
detections can predict mergers detectable by the ground-based detectors,
hundreds of events could also be extracted from the LISA data stream using
the information from ground detection as a prior \citep{Gerosa:2019dbe}.

However, \citet{Moore:2019pke} pointed out that, because of template bank
searching, in order to draw a confident detection, BBH signals in LISA
require a larger signal-to-noise ratio (SNR) threshold than $\rho\sim15$.
The threshold could be reduced to $\rho\sim9$ in combination with
ground-based observation. Even without the limitation of template bank
searching, the SNR threshold could only be lowered to $\rho\sim5$ in the
retracing mode \citep{Wong:2018uwb}. For a GW150914-like source in LISA
band, \citet{Cornish:2018dyw} have calculated its SNR to be $\simeq$ 4 for
a 4-year observation, marginally satisfying the criterion. Nevertheless,
such a BBH source will have a very large SNR if it is seen in the decihertz
band \citep{Isoyama:2018rjb}. Therefore, space-borne detectors of decihertz
band will be extremely helpful in achieving multiband observation, as they
will have larger SNRs and hence better estimated parameters for
stellar-mass BBHs.

\citet{Sedda:2019uro} have proposed Decihertz Observatories (DOs) which are
sensitive in the frequency range of 0.01--1\,Hz. DOs have two LISA-like
illustrative designs, the more ambitious DO-Optimal and the less
challenging DO-Conservative. They can be two representatives of the
space-borne detectors in this frequency range. DECihertz laser
Interferometer Gravitational wave Observatory
\citep[DECIGO;][]{Yagi:2011wg,Kawamura:2018esd} is a more ambitious design
proposed by Japanese scientists. It uses Fabry-Perot cavities as
interferometers and has a lower noise power. The complete design consists
of four independent constellations, which further improves the sensitivity
in the decihertz band and DECIGO's ability in source localization.

For ground-based detectors in the near future, the current LIGO is
gradually upgrading to the full Advanced LIGO (AdvLIGO) at designed
sensitivity and further to A+LIGO \citep{Shoemaker:2010misc}. The
third-generation detectors, Cosmic Explorer \citep[CE;][]{Evans:2016mbw}
and Einstein Telescope \citep[ET;][]{Hild:2010id} are under active
investigation respectively in the United States and Europe. They are
expected to come into use around 2030s.

Future GW detectors on ground and in space call for a strong promise for
multiband observation, thus enabling unprecedented science in the field
\citep{Cutler:2019krq,Barausse:2020rsu}. In this work we study the
multiband prospects for the LIGO/Virgo BBHs in the GW Transient Catalog
\citep[GWTC-1;][]{LIGOScientific:2018mvr} from the LIGO/Virgo's first and
second observing runs.

Multiband observation has been investigated for parameter estimation of BBH
and BH--Neutron Star (NS) systems in general relativity (GR).
\citet{Isoyama:2018rjb} used non-precessing restricted post-Newtonian (PN)
waveform including tidal correction for NSs to estimate parameter precision
of binary systems observed by the baseline DECIGO (B-DECIGO) alone, or
B-DECIGO in combination with ground-based detectors. {\citet{Nair:2015bga}
and \citet{Nair:2018bxj} studied the synergy between the evolved LISA
(eLISA) or DECIGO with ground-based detectors. They respectively studied
cases with and without the information of binary systems' sky location.
Constraints on parameters from joint observation were also examined in e.g.
\citet{Vitale:2016rfr} and \citet{Vitale:2018nif}.

It was pointed out that, multiband observation could also yield stringent
tests of GR \citep{Barausse:2016eii}. Parameterized tests which focus on
the non-GR deviation at different PN orders \citep{Gnocchi:2019jzp,
Carson:2019rda, Toubiana:2020vtf} and inspiral-merger-ringdown consistency
tests which compare the results derived from the inspiral portion and the
merger-ringdown portion separately \citep{Vitale:2016rfr, Carson:2019kkh}
have been carried out. In addition, for the ringdown signal alone,
\citet{Tso:2018pdv} calculated the improvement brought by LISA after it has
informed LIGO ringdown observation with narrow-band tunings.

Soon after the detection of the first BBH event, \citet{Gaebel:2017zys}
studied in detail the signal characteristics for GW150914 in future GW
detector network. Now, the catalog GWTC-1 \citep{LIGOScientific:2018mvr}
has provided us all the information of the confident sources observed
during the first and second observing runs of LIGO and Virgo. Including
GW150914, there are ten BBH events in the catalog and they represent the
most likely observed population of the sources for the future. Therefore,
we use them as indubitable examples and are interested in how they will
look like in future multiband observation. We hope such a study, augmenting the 
existing investigations, can help to extract more clues for scientific
goals.

In this paper, we investigate the projected result of LIGO/Virgo BBH
systems in future GW detectors. We choose eight different GW detectors
across the GW waveband from millihertz to kilohertz. Using the
Fisher matrix analysis method \citep{Finn:1992wt, Cutler:1994ys}, we
estimate the parameter precision in different observational scenarios
including space-based observation, ground-based observation and multiband
observation. Following this, we further perform the Fisher matrix analysis
to a parameterized GW dipole radiation and provide the multiband
constraints on it. In this work, we only consider circular BBHs. The
inclusion of eccentricity has interesting potential to provide
astrophysical information, in particular concerning the formation mechanism
of BBHs, either from galactic fields via traditional binary evolution or in
dense nucleus clusters via dynamical exchange processes
\citep{Nishizawa:2016jji, Chen:2017gfm, Zhang:2019puc, Gerosa:2019dbe,
Liu:2019jpg}. It deserves further investigation.

Multiband parameter estimation for DECIGO in previous work
\citep{Nair:2015bga, Nair:2018bxj, Isoyama:2018rjb} used inspiral-only PN
waveform as the waveform model. When the BBHs are approaching closer and
closer, the PN-only waveform might deviate from the true GR waveform in the
late inspiral and merger-ringdown stages. For BBH systems, such deviation
is relevant to DECIGO or ground-based detectors, hence the inclusion of the
complete inspiral-merger-ringdown waveform will give more realistic
results. Therefore in our work we use the phenomenological waveform model,
the so-called IMRPhenomD waveform \citep{Khan:2015jqa,Husa:2015iqa}. In
addition, different from the single-constellation DECIGO these authors have
considered, we choose the final design of DECIGO that consists of four
constellations, which allows us to study the effect of detectors' orbital
configuration, in particular showing its ability in source localization
even when the observational span is relatively short. On the other hand,
multiband constraints on dipole radiation carried out by previous authors
\citep{Barausse:2016eii, Gnocchi:2019jzp, Carson:2019kkh} used the
sky-averaged waveform, thus ignoring the waveform modulation caused by the
orbital motion of the space-borne detectors. Here, we take into account the
orbital effects and provide additional parameter estimation on the source
location in our Fisher matrix analysis. In contrast to generate a population of BBHs
\citep{Vitale:2016rfr} or using only GW150914-like events
\citep{Carson:2019rda}, our work includes all the LIGO/Virgo BBH events in
GWTC-1 as a population of realistic examples.

The organization of the paper is as follows. In Section~\ref{sec:bbh} we
discuss the GW waveform model, detectors' configuration and detectability,
and the SNRs of BBH sources in different detectors. In
Section~\ref{sec:pe}, we review the Fisher matrix analysis technique and
display our results of source parameter estimation for ground-based,
space-based, and joint observations. In Section~\ref{sec:dipole} we
describe the parameterized post-Einsteinian (ppE) formalism in the presence
of dipole radiation, and perform parameter-estimation study for various
observation scenarios. In Section~\ref{sec:disc} we briefly summarize our
work. Throughout this paper we use geometrized units in which $G=c=1$.

\section{Multiband BBH signals}
\label{sec:bbh}

In this section, we will lay out our setup and investigate the prospects in
observing the BBH sources in the GWTC-1 with future GW detectors. We use
parameters of the ten BBH sources obtained from the Bayesian inference by
the LIGO/Virgo Collaboration \citep{LIGOScientific:2018mvr}. We take the
median value of the source frame component masses $m^{\rm src}_1$ and
$m^{\rm src}_{2}$, redshift $z$, luminosity distance $D_L$, and the maximum
probabilistic sky location $\left(\bar\theta_S, \bar\phi_S \right)$, from
the public posteriors for our calculation. Note that we use the convention,
$m^{\rm src}_1 \geq m^{\rm src}_2$. Because of the cosmological expansion,
the observed waveform is characterized by the redshifted mass {$m_{1,2} =
(1+z)\,m_{1,2}^{\rm src}$}. Therefore we use $m_1, m_2$ in the waveform
model. Since the dimensionless component spins, $\chi_1$ and $\chi_2$,
provided by the GWTC-1 have relatively broad distributions and most of them
are consistent with zero, we choose $\chi_{\rm 1} = \chi_{\rm 2} = 0$ as
the fiducial component spin parameters for each BBH source. Other choices
are unlikely to affect our result in a significant way. We list some of the
source parameters~\citep{LIGOScientific:2018mvr} in Table~\ref{tab:snr} for
readers' convenience.

\subsection{Waveform}\label{sec:waveform}

\begin{figure}
	\includegraphics[width=8.5cm]{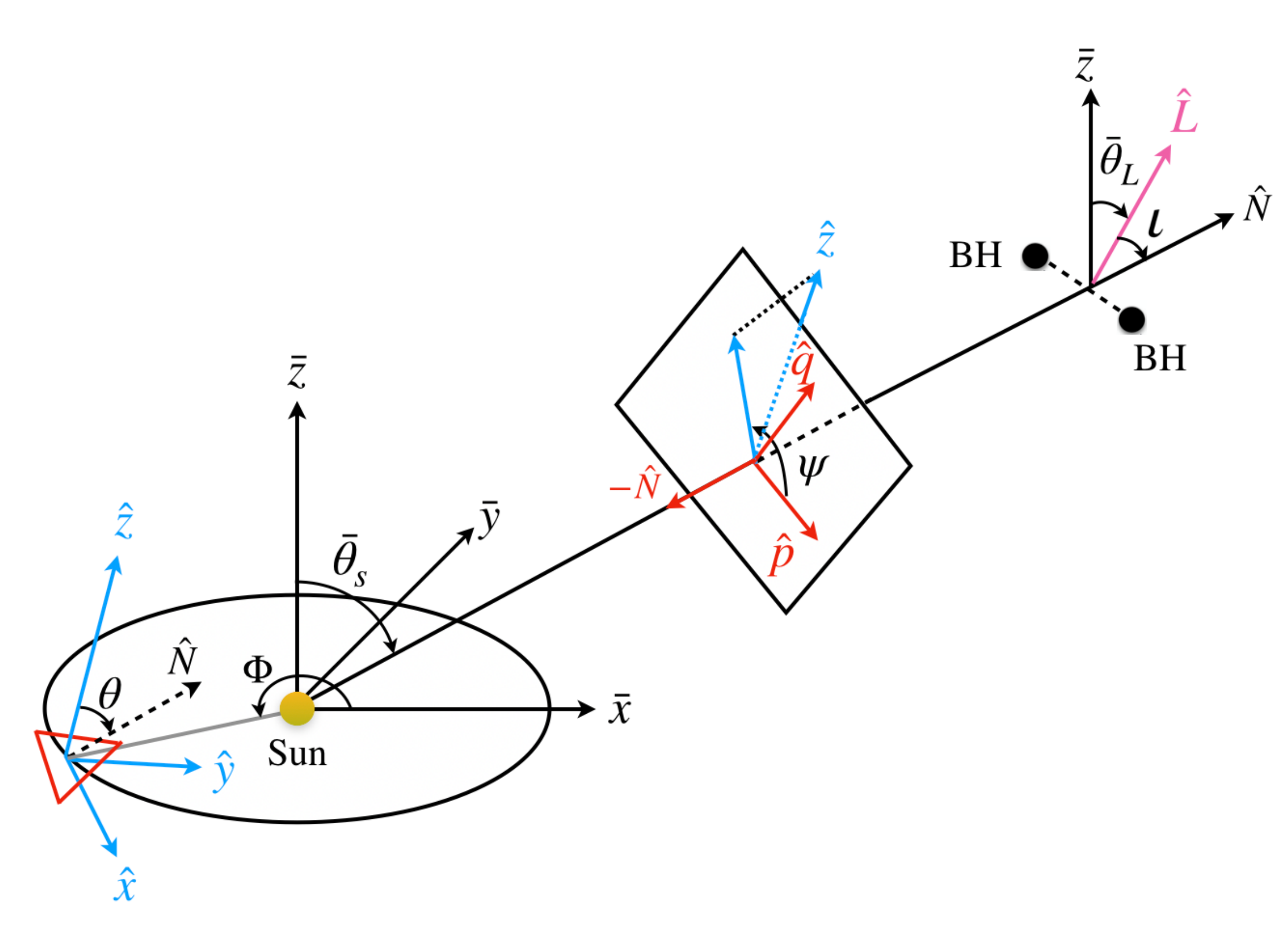}
	\caption{Illustration of various notations that characterize the
	emission, propagation and detection of GWs. Cartesian coordinate system
	($\bar{x} ,\bar y, \bar z$) is attached to the ecliptic. ``Hatted''
	symbols are the unit vectors defined in Section~\ref{sec:waveform}. The
	definitions of the polar angles $\theta$, $\bar\theta_S$,
	$\bar\theta_L$, the polarization angle $\psi$, the inclination angle
	$\iota$, and the azimuthal angle of the detector $\Phi$, are also given
	in Section~\ref{sec:waveform}. Notice that $\theta$ and $\bar\theta_S$
	are the polar angles of the source observed in the detector frame and
	ecliptic frame respectively. }
    \label{fig:angle}
\end{figure}

The GW signal received by the detector is a result of the source waveform
projected on the observers' frame. The projection is described by
direction-dependent pattern functions. In this paper, we use the spinning,
non-precessing IMRPhenomD model \citep{Husa:2015iqa, Khan:2015jqa} as the
source waveform. IMRPhenomD is an inspiral-merger-ringdown waveform for
BBHs, calibrated to the hybrid waveform of numerical relativity and
uncalibrated spinning effective-one-body model \citep{Buonanno:1998gg,
Taracchini:2012ig, Bohe:2016gbl}. It provides us the $l=2,\ |m|=2$
spin-weighted spherical-harmonic modes of the GW. The waveform is a
function of the following physical parameters,
\begin{equation}
	\label{eq:basic}
	\bm{\Xi}^{(0)} = \big\{ D_L, {\cal M}, \eta, t_c, \phi_c,
	\chi_1, \chi_2 \big\} \,,
\end{equation}
where $D_L$ is the luminosity distance of the source; ${\cal M}=M\eta
^{3/5}$ is the chirp mass with the total mass $M=m_1 + m_2$ and the
symmetric mass ratio $\eta=m_1m_2/M^2$; $t_c$ and $\phi_c$ are respectively
the time and phase at coalescence; $\chi_1$ and $\chi_2$ are the
dimensionless spins of BHs.

The frequency-domain GW signal in the detector, $\tilde{h}(f)$, is related
to the incident cross and plus GW signals via
\begin{equation}\label{eq:hf}
	\tilde{h}(f)=F^{+}(f) \tilde{h}_{+}(f)+F^{ \times}(f) \tilde{h}_{
	\times}(f)\,,
\end{equation}
where $\tilde{h}_{+}(f)$ and $\tilde{h}_{ \times}(f)$ are the source GW
waveform provided by IMRPhenomD; $F^{+}(f) $ and $F^{+}(f) $ are
frequency-dependent detector pattern functions. Notice that
Eq.~(\ref{eq:hf}) is valid for space-based detections under the low
frequency approximation, while BBH signals are at the high frequency region
in the LISA band. Therefore the full LISA response should be considered
instead \citep{Marsat:2020rtl}. However, as we will see, the SNRs of BBH
signals in LISA are small, thus the deviation of low frequency
approximation from the full response only has marginal effects
\citep{Rubbo:2003ap}. For computational simplicity, we adopt the low
frequency approximation for all the space-based detectors we considered
(including LISA) in our study.

The pattern functions depend on the sky location ($\theta,\,\phi$) and the
polarization angle $\psi$ of the source (see Fig.~\ref{fig:angle} for
notations). In general, the effect of the inclination angle $\iota$, which
is the angle between the line of sight and the BBH's orbital angular
momentum, is included in the source waveform, since $\iota$ changes with
the precession of the binary's orbital angular momentum. However, because
we consider the non-precessing waveform IMRPhenomD, the inclination angle
$\iota$ is constant. Therefore we take a choice to put $\iota$ in the two
pattern functions,
\begin{align}
	F^{+}(\theta, \phi, \psi,\iota)  
		 &=\frac{\big(1+\cos ^{2} \iota\big)}{2}\ \left[\frac{1}{2}
		 \big(1+\cos ^{2} \theta \big) \cos 2 \phi \cos 2 \psi \right.
		 \nonumber \\
		 & \hspace{2.2cm} -\cos \theta \sin 2 \phi \sin 2 \psi \bigg] \,,
		 \label{eq:Fplus} \\
		 F^{\times}(\theta, \phi, \psi,\iota)
		 &=\cos \iota\ \left[ \frac{1}{2} \big(1+\cos ^{2} \theta \big)
		 \cos 2 \phi \sin 2 \psi \right. \nonumber \\
		 & \hspace{1.2cm} +\cos \theta \sin 2 \phi \cos 2 \psi \bigg] \,.
		 \label{eq:Fcross}
\end{align}
After we include the inclination angle in the pattern functions,
$\tilde{h}_{+}(f)$ and $\tilde{h}_{ \times}(f)$ will have the same
amplitude with a $90^\circ$ phase difference.

Then we consider the averages of pattern functions over different angles,
which are useful when we do not care about the location or the orientation
of the source. The sky-polarization (described by $\theta$, $\phi$, and
$\psi$) averaged function for a quantity $X$ is defined as
\begin{equation}\label{eq:average:3angle}
	\begin{aligned}
	\Big\langle X \Big\rangle_{(3)} \equiv \frac{1}{4 \pi^{2}} &\int_{0}^{\pi} \mathrm{d}
	\psi \int_{0}^{2 \pi} \mathrm{d} \phi \int_{0}^{\pi} X \sin
	\theta\,\mathrm{d} \theta \,,
	\end{aligned}
\end{equation}
where the subscript ``(3)'' denotes that we are averaging over three angles
($\theta$, $\phi$, and $\psi$). Therefore, from Eq.~(\ref{eq:Fplus}) and
Eq.~(\ref{eq:Fcross}) we have
\begin{equation}
    \left\langle F_{+}^2(\theta, \phi, \psi, \iota=0) \right\rangle_{(3)} = 
	\left\langle F_{\times}^2(\theta, \phi, \psi, \iota=0) \right\rangle_{(3)} = \frac{1}{5}\,.
	\label{eq:3angles}
\end{equation}
If we further include the average over the inclination $\iota$,
Eq.~(\ref{eq:average:3angle}) becomes
\begin{equation}
	\begin{aligned}
		\Big\langle X \Big\rangle_{(4)} \equiv \frac{1}{8 \pi^{2}}
		&\int_{0}^{\pi} \mathrm{d} \psi \int_{0}^{2 \pi} \mathrm{d} \phi
		\int_{0}^{\pi} \sin \theta\,\mathrm{d} \theta \int_{0}^{\pi} X \sin
		\iota \, \mathrm{d} \iota \,,
	\end{aligned}
\end{equation}
where the subscript ``(4)'' denotes that we are averaging over four angles
($\theta$, $\phi$, $\psi$, and $\iota$). Now we have
sky-polarization-inclination averaged pattern functions,
\begin{equation}
	\label{eq:ground12}
 \left\langle F_{+}^2(\theta, \phi, \psi,\iota) \right\rangle_{(4)} =
 \frac{7}{75} \,, \quad 
\left\langle F_{\times}^2(\theta, \phi, \psi,\iota) \right\rangle_{(4)} =
\frac{1}{15} \,.
\end{equation} 

The above formulae apply to both ground-based and space-based detectors.
For sensitivity curves plotted in {Fig.}~\ref{fig:gw:source} and SNR
calculations in Table~\ref{tab:snr}, we ignore the difference caused by
source's location and orientation. Thus we have used
\begin{equation}
	\label{eq:average}
	\Big\langle F_{+}^2(\theta, \phi, \psi,\iota) \Big\rangle_{(4)} +
	\Big\langle F_{\times}^2(\theta, \phi, \psi,\iota) \Big\rangle_{(4)} =
	\frac{4}{25} \,,
\end{equation} 
as the average term and will explain in detail in Section~\ref{sec:snr}.
Next, we discuss separately how we deal with these two classes of GW
detectors when performing the parameter estimation.

For ground-based detectors, we take the sky-averaged waveform. The reason
is that, usually the localization power of ground-based detectors do not
depend on their ability to distinguish $\theta$, $\phi$, $\psi$, and
$\iota$, but rather depend on the timing measurement between individual
detectors \citep{Fairhurst:2010is}. We use the square root of
Eq.~(\ref{eq:ground12}) in parameter estimation for ground-based detectors.

For space-based detectors, since their localization ability depends on the
direct measurement of $\theta$, $\phi$, $\psi$, and $\iota$
\citep{Cutler:1997ta}, we take the non-sky-averaged waveform in the
parameter estimation. Because of their orbital motion during the
observation, $\theta$, $\phi$, and $\psi$ are functions of time $t$. To
estimate the localization power, we describe the motion of the detectors
using the ecliptic coordinate ($\bar x, \bar y, \bar z$). Following the
method in \citet{Cutler:1997ta}, we re-express $\big\{\theta, \phi,
\psi,\iota\big\}$ by the time variable $t$ and
\begin{equation} \label{eq:Xi:loc}
	\bm{\Xi}^{\rm loc} \equiv \big\{\bar \theta_S, \bar\phi_S,\bar\theta_L,
	\bar\phi_L\big\} \,.
\end{equation}
Specifically, we have
\begin{align}
	\cos \theta &= \hat{N} \cdot \hat{z}\,, \label{eq:space:costheta}\\
	\label{eq:alpha0}
	\phi &= \arctan \left(\frac{\hat{N} \cdot \hat{y}}{\hat{N} \cdot
	\hat{x}}\right) + \frac{2\pi t}{T} + \alpha_0\,,\\
	\tan \psi &=  \frac{\hat{z} \cdot \hat{q} }{\hat{z} \cdot
	\hat{p}}\,,\\
	\cos \iota &= \hat{N} \cdot \hat{L}\,, \label{eq:space:cosiota}
\end{align}
where $T$ is the orbital period of the detector, which equals to one year
for heliocentric orbit, and the angle $\alpha_0$ is the initial orientation
of the detector arms. We illustrate various vectors appearing in
Eqs.~(\ref{eq:space:costheta}--\ref{eq:space:cosiota}) in
Fig.~\ref{fig:angle}, and explain them in the following. 

The unit vector $\hat{L}$ is parallel to the orbital angular momentum of
the BBH, while expressed in the ecliptic coordinate,
\begin{equation}
	\hat{L} = \left(\sin\bar\theta_L
	\cos\bar\phi_L,\,\sin\bar\theta_L\sin\bar\phi_L,\,
	\cos\bar\theta_L\right)\,.
\end{equation}
The unit vector $\hat{N}$ is the line-of-sight vector pointing from the
detector to the binary system, while expressed in the ecliptic coordinate,
\begin{equation}
	\hat{N} = \left(\sin\bar\theta_S\cos\bar\phi_S,\,
	\sin\bar\theta_S\sin\bar\phi_S,\, \cos\bar\theta_S\right)\,.
\end{equation}
The unit vectors $\hat{p}$ and $\hat{q}$ are the axes orthogonal to
$\hat{N}$ with $\hat{p} = \hat{N} \times \hat{L}$ and $\hat{q} = - \hat{N}
\times \hat{p}$. The coordinates $(\hat x, \hat y, \hat z)$ are given by,
\begin{align}
	\hat{x} &= \Big(-\sin\Phi(t),\, \cos\Phi(t),\, 0\Big)\,,\\
	\hat{y} &= \Big(-\frac{1}{2}\cos{\Phi}(t),\,
	-\frac{1}{2}\sin{\Phi}(t),\, -\frac{\sqrt{3}}{2}\Big)\,, \\
	\hat{z} &= \Big(-\frac{\sqrt{3}}{2}\cos{\Phi}(t), \,
	-\frac{\sqrt{3}}{2}\sin{\Phi}(t),\, \frac{1}{2}\Big) 
	\,,
\end{align}
where ${\Phi}(t)$ is the azimuthal angle of the detector around the Sun,
\begin{equation}
	{\Phi}(t) = \Phi_0 + \frac{2\pi t}{T}\,,
	\label{eq:phi0}
\end{equation}
with $\Phi_0$ an initial angle at $t=0$. In this way, $\hat{z}$ is the
normal vector to the detector plane, which has a constant inclination,
60$^\circ$, to the $\bar z$ axis. Note that the unit vector $\hat{x}$ is
always in the ecliptic plane and $\hat{y} = \hat{z} \times \hat{x}$. The
$\hat{x}$ and $\hat{y}$ we defined here do not rotate with the detector,
while the term ${2\pi t}/{T}$ in {Eq.~(\ref{eq:alpha0}) comes from the
self-rotation of the detector. Figure~\ref{fig:angle} shows the Cartesian
coordinate ($\bar x,\bar y,\bar z$) tied to the ecliptic, which
characterizes the polar angles ($\bar \theta_S, \bar\theta_L$). It also
shows the aforementioned angles ($\theta, \psi, \iota$), as well as the
unit vectors that are defined above.

Because of the orbital motion of space-based detectors around the Sun,
there is a Doppler phase correction \citep{Cutler:1997ta},
\begin{equation}
	\varphi_{D}(t)={2 \pi f(t)}\, R \sin {\bar\theta}_{S} \cos
	\left[{\Phi}(t)-\bar\phi_S\right]\,,
\end{equation}
where $R = 1$\,AU is the orbital radius of the detector. For the GW
waveform in time domain, we have
\begin{align}
	{h}(t) &= e^{i{\varphi_D}(t)} \left[ F^{+} \big( \bm{\Xi}^{\rm loc}, t
	\big) \, {h}_{+}(t) +F^{ \times} \big( \bm{\Xi}^{\rm loc}, t \big) \,
	{h}_{ \times}(t)\right] \,.
\end{align} 

Because all the modulations encoded in the pattern functions and the
Doppler phase vary on time scales of 1 year $\gg$ 1/$f$ where $f$ is the GW
frequency, we can approximate $\tilde{h}(f)$ using the stationary phase
approximation \citep[SPA; see e.g.\ ][]{Feng:2019wgq}. Therefore in the
frequency domain, the waveform becomes
\begin{align}
	\tilde{h}(f) &= e^{-i{\varphi_D}(f)} \left[ F^{+}\big( \bm{\Xi}^{\rm
	loc}, f \big) \, \tilde{h}_{+}(f) +F^{ \times} \big( \bm{\Xi}^{\rm
	loc}, f \big) \, \tilde{h}_{ \times}(f)\right] \,.
\end{align}
In the consideration that $\tilde{h}(f)$ is a function of
{$\bm{\Xi}^{(0)}$} and the above equations, we express our final waveform
for space-borne detectors in the full form
\begin{align}
	\tilde{h}(f)&= e^{-i{\varphi_D} \left(\bar\theta_S,\bar\phi_S,f,{\cal
	M},t_c \right)} \Big[ F^{+} \big( \bm{\Xi}^{\rm loc}, f,{\cal M},t_c
	\big) \, \tilde{h}_{+}\big( \bm{\Xi}^{(0)} ,f \big) \nonumber \\
	& \hspace{2.5cm} +\,F^{ \times} \big( \bm{\Xi}^{\rm loc}, f,{\cal
	M},t_c \big) \, \tilde{h}_{ \times} \big(\bm{\Xi}^{(0)},f \big) \Big]
	\,. \label{eq:waveform}
\end{align}
Note that our waveform construction will be the same as that of
\citet{Cutler:1997ta} if we change our inspiral-merger-ringdown waveform to
the restricted 1.5\,PN inspiral waveform.

\subsection{Detectors}
\label{sec:detector}

\begin{figure*}
	\includegraphics[width=17cm]{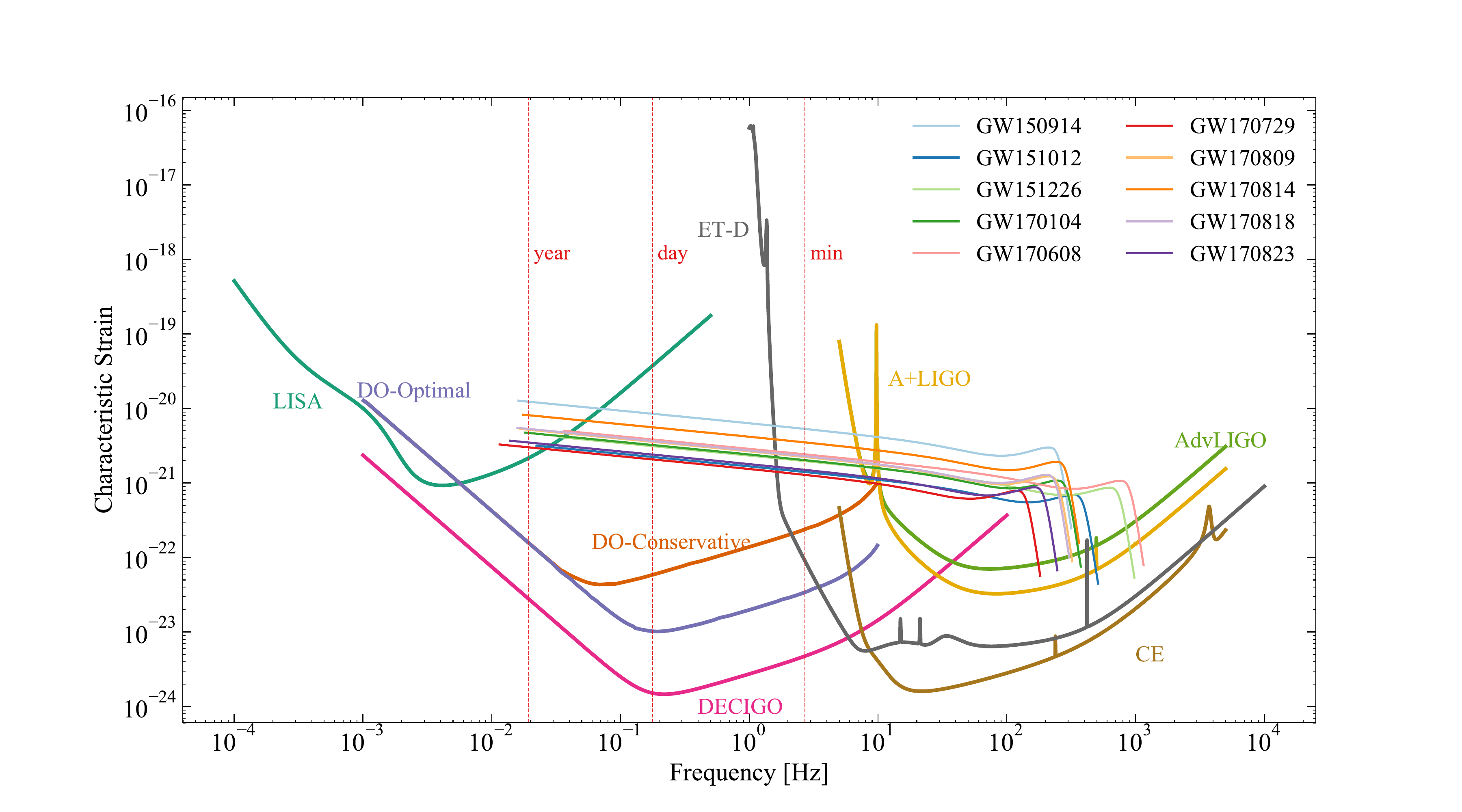}
    \caption{Multiband observation of LIGO/Virgo BBHs
    \citep{LIGOScientific:2018mvr} with ground-based and space-based GW
    detectors. The effective strain amplitude of the source, ${2f| \tilde
    h_+(f)|}$, is plotted against the characteristic strain of the
    effective detector noise, $\sqrt{fS_{n}^{\rm eff}(f)}$. BBH signals are
    plotted for a duration of $T_{\rm obs} = 4$\,yrs. Dashed lines mark the
    coalescence times for the most massive system in the GWTC-1, GW170729
    \citep{Chatziioannou:2019dsz}. }
    \label{fig:gw:source}
\end{figure*}

\begin{table}
	\centering
	\caption{The equivalently independent number of detectors, geometrical
	configuration, and frequency range of future GW detectors that we use
	in this work. }
	\label{tab:detector}
	\def\arraystretch{1.3}
	\begin{tabular}{lclrl} 
		\hline\hline
		Detector &Number &Configuration & $f_{\rm low}$(Hz) & $f_{\rm high}$(Hz) \\
		\hline
		LISA&2&Triangle&0.0001&0.5\\
		DO &2&Triangle&0.001&10 \\
		DECIGO &8&Triangle&0.001&100 \\
		AdvLIGO &2&Right-angle&5&5000 \\
		A+LIGO &2&Right-angle&5&5000 \\
		CE &2&Right-angle&5&5000 \\
		ET &3&Triangle&1&10000 \\
		\hline
	\end{tabular}
\end{table}

From the proposed GW detectors across different wavebands, we choose these
representatives in our study:
\begin{enumerate}[(i)]
	\item LISA in millihertz band;
	\item DO-Conservative, DO-Optimal and DECIGO in decihertz band;
	\item AdvLIGO, A+LIGO, CE and ET in hectohertz band.
\end{enumerate}
We list the equivalent number of detectors, their geometrical configuration
and frequency range in Table~\ref{tab:detector}. Below we list the relevant
literature of our sensitivity curves and the way to convert them equivalent
to a non-sky-averaged single-detector noise power density $S_{n}(f)$.
\begin{itemize}
  \item For LISA, we take Eq.~(1) in \citet{Cornish:2018dyw} and because it
  is already a $(\theta,\phi,\psi)$-averaged noise power spectral density
  of the 2-channel (equivalent to 2 effective detectors), triangle-shape
  detector, we further times $2/5 = 2 \times 1/5$ [see
  Eq.~(\ref{eq:3angles})] to get the non-sky-averaged single detector
  noise.
  \item For DO-Optimal and DO-Conservative, we take the sensitivity curves
  from \citet{Sedda:2019uro} and deal with them in the same way as we deal
  with LISA's curve.
  \item For DECIGO, we take the L-shape power spectral density from
  \citet{Yagi:2011wg} and times ${\big(\sin ^2 60^{\circ}\big)}^{-1} $ to
  convert it to the triangle-shape detector.
  \item For AdvLIGO and A+LIGO, we take the power spectral density from
  LIGO documents.\footnote{\url{https://dcc.ligo.org/LIGO-T1800044/public}
  and \url{https://dcc.ligo.org/LIGO-T1800042/public}}
  \item For CE, we take the power spectral density from
  \citet{Evans:2016mbw}, and learn from \citet{Reitze:2019dyk} that two CE
  detectors are envisioned as part of the ground-based network.
  \item For ET, we choose the final ET-D design. We take the L-shape ET-D
  power spectral density from \citet{Hild:2010id} and times ${\big(\sin ^2
  60^{\circ}\big)}^{-1} $ to convert it to the triangle-shape power
  spectral density.
\end{itemize}
The non-sky-averaged single-detector noise power density $S_{n}(f)$ that we
collected above is used in the parameter estimation in Section~\ref{sec:pe}
and Section~\ref{sec:dipole}, while for the SNR calculation in
Section~\ref{sec:snr}, we instead use an effective noise power $S_{n}^{\rm
eff}(f)$ [see Eq.~(\ref{eq:effective})]. The effective noise curves are
plotted for the above detectors in Fig.~\ref{fig:gw:source}.

\subsection{SNRs} \label{sec:snr}

\def\arraystretch{1.3}
\begin{table*}
	\centering
	\caption{Some source parameters and the
	$\left(\theta,\phi,\psi,\iota\right)$-averaged SNRs for LIGO/Virgo BBHs
	in different GW detectors.The listed BBH parameters are the
	source-frame component masses $m_1^{\rm src}$ and $m_2^{\rm src}$, the
	chirp mass ${\cal M}^{\rm src}$, and the luminosity distance $D_L$. We
	list the median values of these parameters with 90\% credible intervals
	\citep{LIGOScientific:2018mvr}. For space-borne detectors, we have
	assumed signals of a 4-year duration.}
	\label{tab:snr}
	\begin{tabular}{l|rrrr|llllllll} 
		\hline\hline
		GW &\multicolumn{4}{c}{Source Parameters}& LISA & \multicolumn{2}{c}{DO} & DECIGO & AdvLIGO & A+ & CE & ET\\
		&$m_1^{\rm src}/M_\odot$&$m_2^{\rm src}/M_\odot$&${\cal M}^{\rm
		src}/M_\odot$&$D_L/\rm Mpc$ && CON & OPT & \\
		\hline
		GW150914 & $35.6^{+4.7}_{-3.1}$ & $30.6^{+3.0}_{-4.4}$ & $28.6^{+1.7}_{-1.5}$ & $440^{+150}_{-170}$ & 
		4.5  &300 & 1100 & 7400  & 52 &110 &2600&960 \\
		GW151012 & $23.2^{+14.9}_{-5.5}$ & $13.6^{+4.1}_{-4.8}$ & $15.2^{+2.1}_{-1.2}$ & $1080^{+550}_{-490}$ & 
		0.81 & 78 &290   & 1900   &14  &29  &700 & 260\\
		GW151226 & $13.7^{+8.8}_{-3.2}$ & $7.7^{+2.2}_{-2.5}$ & $8.9^{+0.3}_{-0.3}$ & $450^{+180}_{-190}$ & 
		0.65 & 110 &400  &2700  &21  &43  &1000 &370 \\
		GW170104 & $30.8^{+7.3}_{-5.6}$ & $20.0^{+4.9}_{-4.6}$ & $21.4^{+2.2}_{-1.8}$ & $990^{+440}_{-430}$ & 
		1.5 & 110 &420  &2800  &20  &42  &990 &370 \\
		GW170608 & $11.0^{+5.5}_{-1.7}$ & $7.6^{+1.4}_{-2.2}$ & $7.9^{+0.2}_{-0.2}$ & $320^{+120}_{-110}$ & 
		0.70 & 130 &490  &3300   &26  &54  &1300 &460 \\
		GW170729 & $50.2^{+16.2}_{-10.2}$ & $34.0^{+9.1}_{-10.1}$ & $35.4^{+6.5}_{-4.8}$ & $2840^{+1400}_{-1360}$ & 
		1.6 & 74 &270   &1800    &12 &25 &600 &220 \\
		GW170809 & $35.0^{+8.3}_{-5.9}$ & $23.8^{+5.1}_{-5.2}$ & $24.9^{+2.1}_{-1.7}$ & $1030^{+320}_{-390}$ & 
		1.9  & 130  &460  &3100   &22 &46  &1100 &400 \\
		GW170814 & $30.6^{+5.6}_{-3.0}$ & $25.2^{+2.8}_{-4.0}$ & $24.1^{+1.4}_{-1.1}$ & $600^{+150}_{-220}$ & 
		2.7  & 200  &720  &4900   &35 &73  &1700 &630 \\
		GW170818 & $35.4^{+7.5}_{-4.7}$ & $26.7^{+4.3}_{-5.2}$ & $26.5^{+2.1}_{-1.7}$ & $1060^{+420}_{-380}$ & 
		2.0  & 130 &470  &3200   &22 &47  &1100 &410 \\
		GW170823 & $39.5^{+11.2}_{-6.7}$ & $29.0^{+6.7}_{-7.8}$ & $29.2^{+4.6}_{-3.6}$ & $1940^{+970}_{-900}$ & 
		1.5 & 85 & 310    &2100   &15 &30 &720 &270 \\
		\hline
	\end{tabular}
\end{table*}

To estimate the SNRs as well as the source parameters in
Section~\ref{sec:pe} and Section~\ref{sec:dipole}, we use the matched
filtering method \citep{Finn:1992wt}. The noise weighted inner product
between two signals $h_1(t)$ and $h_2(t)$ is defined as
\begin{equation}
	(h_1|h_2) \equiv 2 \int_{f_{\rm lower}}^{f_{\rm upper}} \frac{
	{\tilde{h}^*_1 (f) } \tilde{h}_2^{} (f) + {\tilde{h}_2}^* (f)
	\tilde{h}_1^{} (f) } {S_n(f)} {\rm d} f \,, \label{innerproduct}
\end{equation}
where $\tilde{h}_1 (f)$ and $\tilde{h}_2 (f)$ are the Fourier transform of
$h_1(t)$ and $h_2(t)$, and $S_{n}(f)$ is the power spectral density of the
detector noise. The SNR for a given signal $h$ is given by
\begin{equation}
	\rho \equiv \sqrt{(h | h)}\,. 
	\label{eq:snr}
\end{equation}
We choose the integral range of frequency in Eq.~(\ref{innerproduct}) as
follows. For ground-based detectors, ${f_{\rm lower}}$ and ${f_{\rm
upper}}$ are the lower cut-off and higher cut-off frequency of the detector
(see Table~\ref{tab:detector}). For space-based detectors, ${f_{\rm
upper}}$ is the higher cut-off of the detector and ${f_{\rm lower}}$ is
given by
\begin{equation}
	f_{\rm lower} = \frac{1}{8 \pi}\left(\frac{ T_{\rm obs} +
	t_{\rm upper}}{5}\right)^{-3/8}{\cal M}^{-5/8}\,,
\end{equation}
where
\begin{equation}
	t_{\rm upper} = 5 \left( 8 \pi f_{\rm upper} \right)^{-8 / 3}\mathcal{M}^{-5 / 3}\,,
\end{equation}
and $T_{\rm obs}$ is the duration of the mission. Such a $f_{\rm lower}$
corresponds to the GW frequency when the mission starts, for a GW signal
leaving the sensitivity band when the mission finishes.

For convenience in comparison, we estimate the SNRs using the $(\theta,
\phi, \psi, \iota)$-averaged effective noise spectral density $S_{n}^{\rm
eff}(f)$, which is defined as
\begin{equation}
	S_{n}^{\rm eff}(f)=\frac{S_{n}(f)}{\mathcal{R}(f)}\,,
	\label{eq:effective}
\end{equation}
where ${S_{n}(f)}$ is the non-sky-averaged single-detector noise power
spectral density defined in Section~\ref{sec:detector} that includes the
effect of the detector shape (i.e. the $\sin ^2 60^{\circ}$ factor if the
detector is triangular); ${\mathcal{R}(f)}$ is the signal response function
of the instrument, which further includes the effect of detector number and
the averages over $(\theta, \phi, \psi, \iota)$ angles. For example,
${\mathcal{R}(f)}$ of LISA is $8/25 = 2 \times 4/25 $, whereas 2 stands for
2 detectors, and $4/25$ stands for the average over 4 angles [see
Eq.~(\ref{eq:average})]. In the same way, ${\mathcal{R}(f)}$ of DECIGO is
$32/25 = 8 \times 4/25 $. We apply the inverse of this factor to the noise
power ${S_{n}(f)}$ to define the effective noise spectral density
$S_{n}^{\rm eff}(f)$. Since the $(\theta, \phi, \psi, \iota)$-averaged
pattern functions have already been considered in $S_{n}^{\rm eff}(f)$ via
Eq.~(\ref{eq:average}), we take $h=h_+(t)$ in Eq.~(\ref{eq:snr}) to
calculate the $(\theta, \phi, \psi, \iota)$-averaged SNRs. The definition
of ${\mathcal{R}(f)}$ is similar to {Eqs.~(2), (7), and (9)} in
\citet{Cornish:2018dyw}, but we have three differences. First is that we
include the average over 4 angles, ($\phi,\,\theta,\,\psi,\,\iota$),
instead of 3 angles, ($\phi,\,\theta,\,\psi$). Second is that we consider
the detector shape effect in the ${S_{n}(f)}$ term instead of the
${\mathcal{R}(f)}$ term. The third one is that for ground-based detectors,
we consider both the number and the configuration for each detector.

In Fig.~\ref{fig:gw:source} we plot the ten BBH sources in GWTC-1 and all
the detector sensitivity curves. In order to have an intuitive feeling of
the SNRs, we plot the effective strain amplitude, ${2f| \tilde h_+(f)|}$,
against the characteristic strain of the detectors' effective noise,
$\sqrt{fS_{n}^{\rm eff}(f)}$. We also note three typical moments (1\,year,
1\,day, 1\,minute) before the coalescence of the most massive source,
GW170729, which has a redshifted chirp mass of $\sim$ {$53.5\,M_\odot$}
\citep{LIGOScientific:2018mvr, Chatziioannou:2019dsz}.

In Table~\ref{tab:snr} we list the ($\phi, \,\theta, \,\psi,
\,\iota$)-averaged SNRs for each LIGO/Virgo BBH source in each detector. We
found that the SNRs of light sources such as GW170608 and GW151226 in LISA
band are very low that we cannot observe them even in a retracing mode. For
instance, the SNR of GW170608 stays around $\rho\simeq0.7$ even if we
change the observing ending time few years before the merger, which means
that GW170608-like sources will remain undetectable in LISA band as long as
the observing duration is 4 years. However, other decihertz space-borne
detectors in higher frequency bands have SNRs far over the $\rho=15$
threshold \citep{Moore:2019pke} and would have certainly recognized all the
GW signals of LIGO/Virgo BBH systems had they in operation.

With the information of the source location, we also calculate their
non-sky-averaged SNRs by randomly generating 1000 different directions for
the orbital angular momentum. The maximum SNR in LISA is 6.5 for the source
GW150914 with an median value of SNR $\rho \sim 1.6$ among all the 1000
samples. It indicates that we could barely observe the LIGO/Virgo BBH
signals in GWTC-1 in LISA, except for fortunate orientations.

\begin{figure}
	\includegraphics[width=8.5cm]{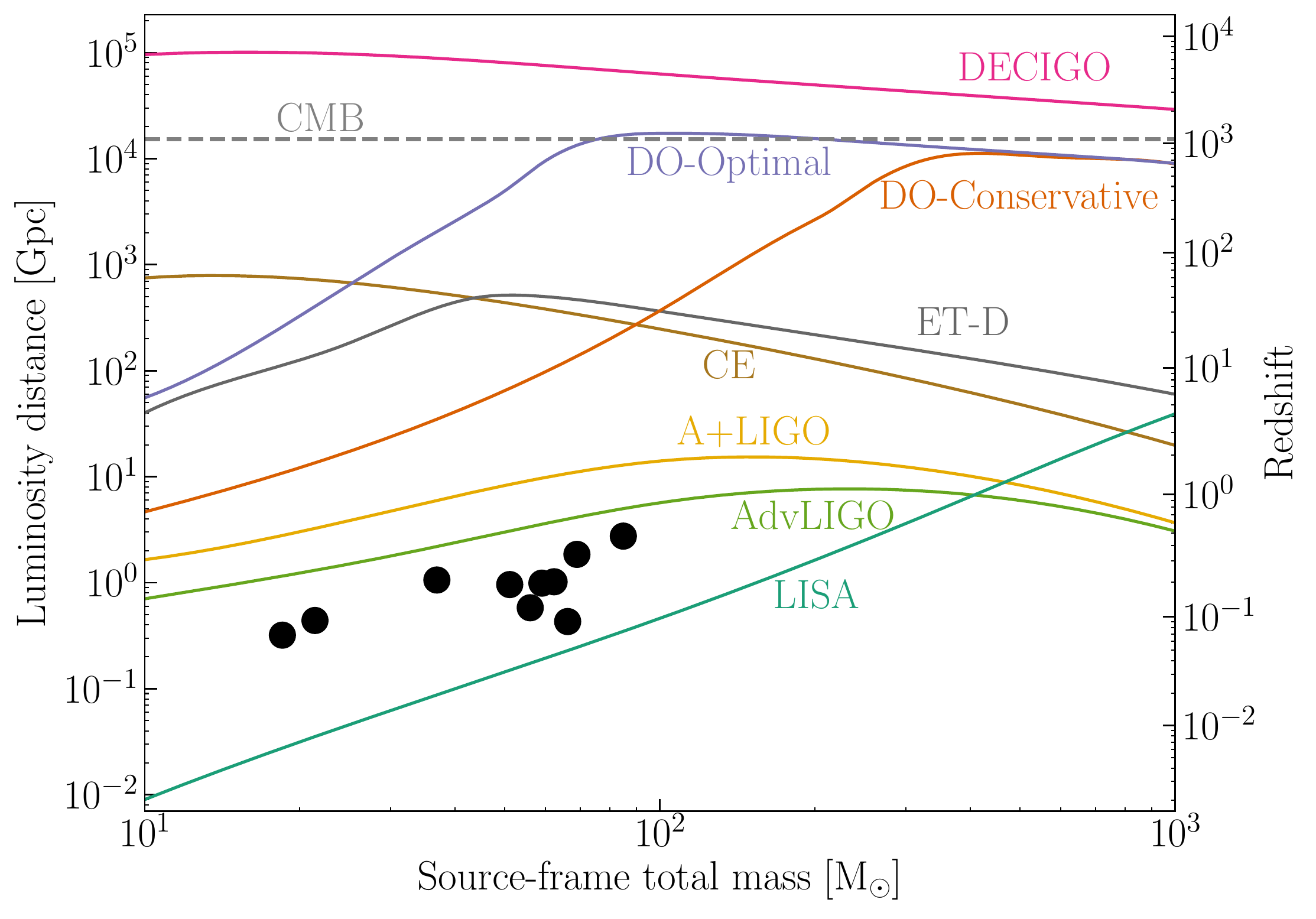}
	\caption{ Cosmological reach to equal-mass BBHs in different detectors.
	The abscissa displays the total mass measured in the source frame.
	Black dots mark the ten GWTC-1 BBHs. The luminosity distance and
	corresponding redshift on the ordinate has been computed at a minimum
	SNR threshold of 8 for all detectors. We assume a 4-year mission
	lifetime for space-borne detectors. For DECIGO we have used the
	sensitivity curve of 8 effective detectors. The dashed line marks the
	time when the recombination began at redshift $z\simeq 1100$. }
    \label{fig:gw:horizon}
\end{figure}

Because of the cosmic expansion, the relation between the luminosity
distance and the redshift $z$ is given by
\begin{equation}
  D_{L}(z)=\frac{1+z}{H_{0}} \int_{0}^{z} \frac{ {\rm d} z^{\prime}}{
  \sqrt{\Omega_{M}\left(1+z^{\prime}\right)^{3}+\Omega_{\Lambda}} }\,.
\end{equation}
In Fig.~\ref{fig:gw:horizon}, we plot the luminosity distance $D_L$ that
could be reached by different detectors. We have used the $\rm \Lambda CDM$
cosmology with matter density parameter $\Omega_M = 0.315$, dark energy
density parameter $\Omega_\Lambda = 0.685$ and the Hubble constant $H_0 =
67.4\, \rm km\, s^{-1}\, Mpc^{-1}$ \citep{Aghanim:2018eyx}. We also adopt
the sky-averaged $S_{n}^{\rm eff}(f)$ in the calculation, with a 4-year
observing time for space-borne detectors. This plot indicates the
detectability of each detector to a large population of stellar-mass BBHs,
with a possible exception of LISA.

\section{Parameter estimation}
\label{sec:pe}

In the matched filtering method, the full Bayesian analysis is usually used
to derive the distribution of source parameters
\citep{LIGOScientific:2018mvr}. We denote source parameters collectively in
$\bm{\Xi}$, with component $\Xi^a$. We define two sets of parameters used
in the following calculation,
\begin{align}
    {\bm{\Xi}^{\rm GR}_{\rm w/o\,loc}} &= \bm{\Xi}^{(0)} \,,
    \label{eq:Xi:noloc} \\
    {\bm{\Xi}^{\rm GR}_{~~~~~~~~~~}} &= \bm{\Xi}^{(0)} \cup \bm{\Xi}^{\rm
    loc} \,, \label{eq:theta:GR}
\end{align}
where $\bm{\Xi}^{(0)}$ and $\bm{\Xi}^{\rm loc}$ are defined respectively in
Eq.~(\ref{eq:basic}) and Eq.~(\ref{eq:Xi:loc}). The set $\bm{\Xi}^{\rm
GR}$ includes the localization parameters, while the set ${\bm{\Xi}^{\rm
GR}_{\rm w/o\,loc}}$ does not.

\subsection{Fisher matrix}

Bayesian analysis could be computationally expansive with a large number of
sources and a high-dimensional parameter space \citep{Toubiana:2020vtf}. In
the limit of large SNRs, if the noise is stationary and Gaussian, the
Fisher matrix method \citep{Finn:1992wt, Cutler:1994ys} is sufficient to
estimate parameter precision instead. Notice that the SNRs of GWTC-1 BBHs
in LISA are small (see Table~\ref{tab:snr}), thus Bayesian analysis is
essential in this situation \citep{Vallisneri:2007ev, Toubiana:2020vtf}.
For detections that have large SNRs, the Fisher matrix method should work
well. Nevertheless, we perform the Fisher matrix to all the detectors, and
we keep caution in interpreting the results related to LISA.

The element of the Fisher matrix $\Gamma_{ab}$ is given by
\citep{Finn:1992wt, Cutler:1994ys}
\begin{equation}
	\Gamma_{ab} \equiv \left( \frac{\partial h}{\partial \Xi^a}
	\right| \left. \frac{\partial h}{\partial\Xi^b} \right)\,.
\end{equation}
The probability that a GW signal $s$ is characterized by source parameters
${\bm \Xi}$ is
\begin{equation}
	p({\bm{\Xi}} | s)=p^{(0)}({\bm{\Xi}}) \exp \left[-\frac{1}{2} \Gamma_{a
	b} \Delta \Xi^{a} \Delta \Xi^{b}\right]\,,
	\label{eq:gauss}
\end{equation}
where $\Delta \Xi^{a} = \Xi^{a} - \hat{\Xi}^{a}$ with $\hat{\Xi}^{a}$ the
maximum-likelihood parameter determined in the matched filtering, and
$p^{(0)}({\bm{\Xi}})$ is a normalization {proportional to} the prior
distribution of the parameter set ${\bm{\Xi}}$. When the prior is absent
(or uniform in $\bm{\Xi}$), Eq.~(\ref{eq:gauss}) is a multivariate Gaussian
distribution and the variance-covariance matrix element is given by
\begin{equation}
\left\langle\delta \Xi^{a} \delta \Xi^{b}\right\rangle=\left(\Gamma^{-1}\right)^{ ab}\,.
\end{equation}
An estimate of the root-mean-square (rms) $\Delta \Xi^{a}$ is then
\begin{equation}
	\Delta \Xi^{a} =\sqrt{ \left(\Gamma^{-1}\right)^{a a}}\,.
\end{equation}

In GR, the dimensionless spins of BHs cannot exceed unity, in order to
satisfy the cosmic censorship conjecture. Like in \citet{Berti:2004bd}, we
take into account this prior information to limit the maximum spin. The
absolute value of the dimensionless spin of each BH should be less than
unity, so here we loosely assume a Gaussian prior for $\chi_1$ and $\chi_2$
with a unity spread, i.e.
\begin{equation}
	p^{(0)}(\chi_1, \chi_2) \propto \exp
	\left[-\frac{1}{2}\big(\chi_1^{2}+\chi_2^{2}\big)\right]\,.
\end{equation}
All other parameters have assumed uninformative priors. 

The angular resolution $\Delta \Omega$ is defined as
\citep{Cutler:1997ta, Barack:2003fp}
\begin{equation}
	\Delta \Omega=2 \pi \sqrt{\left(\Delta \bar\mu_{S} \Delta
	\bar\phi_{S}\right)^{2}-\left\langle\delta \bar\mu_{S} \delta
	\bar\phi_{S}\right\rangle^{2}}\,,
\end{equation}
where $\Delta\bar\mu_{S}$ and $ \Delta \bar\phi_{S}$ are the rms errors of
$\bar\mu_{S}$ and $\bar\phi_{S}$ with $\bar\mu_{S} \equiv
\cos\bar\theta_S$, and $\left\langle\delta \bar\mu_{S} \delta
\bar\phi_{S}\right\rangle$ is the covariance of $\bar\mu_{S}$ and
$\bar\phi_{S}$. Note that $\bar \theta_S, \bar\phi_S$ are expressed in the
ecliptic coordinate. The fiducial values of the parameters are discussed at
the beginning of Section~\ref{sec:bbh}; in addition, without losing
generality, we choose {$t_c=\phi_c=0$}.

For a given detector, the Fisher matrix of the detector is $ \Gamma_{ab}^{\rm total} = \sum_{ k = 1}^{ N}
\Gamma_{ab}^{k}$, where $\Gamma_{ab}^{k}$ is the Fisher matrix of the
$k^{\rm th}$ effective detector and $N$ is the total number of the
effective detectors.
To evaluate the Fisher matrix, we need to calculate the partial derivative
of $\tilde{h}(f)$ in Eq.~(\ref{eq:waveform}) with respect to each
parameter. We calculate the partial derivatives of $\tilde{h}(f)$ with
respect to $t_c$ and $\phi_c$ analytically, giving ${\partial
\tilde{h}}/{\partial t_{c}}=2 \pi \mathrm{i} f \tilde{h}$ and ${\partial
\tilde{h}}/{\partial \phi_{c}}=-\mathrm{i} \tilde{h}$; we calculate the
partial derivatives of $\tilde{h}$ with respect to other parameters
numerically. Note that the variation caused by changing a small amount of
$t_c$ and ${\cal M}$ is trivial in $F^{+,\times}$ and $\varphi_D$,
comparing to that in $\tilde{h}_+$ and $\tilde{h}_\times$. For simplicity,
we set $\partial F^{+,\times}/\partial t_c = \partial F^{+,\times}/\partial
{\cal M} = \partial \varphi_D/ \partial t_c = \partial \varphi_D/ \partial
{\cal M} = 0$; we have checked that it only affects our results at the
relative level of $\sim 10^{-5}$.

\subsection{Parameter estimation with space-borne detectors}

In low frequency approximation, a triangular space-based interferometer can
be seen as two independent detectors, in which one rotates by $45^\circ$
with respect to the other one \citep{Cutler:1997ta}, thus the angle
$\alpha_0$ in Eq.~(\ref{eq:alpha0}) equals $0^\circ$ and $45^\circ$
respectively for the two independent detectors. This is the case with LISA
and DOs. In the same way, four triangular DECIGO constellations can be seen
as eight effective detectors. In addition, these eight detectors are
divided into three groups with detector numbers $(2,2,4)$, located from one
another by $120^\circ$ on their heliocentric orbits \citep{Yagi:2011wg}.
Thus we set respectively the initial angle $\Phi_0$ in Eq.~(\ref{eq:phi0})
of the three groups $0^\circ$, $120^\circ$, and $240^\circ$
($\Phi_0=0^\circ$ for LISA and DOs). {As we will see, the full
configuration of DECIGO will} have a huge advantage in sky localization.

For the parameter estimation in GR, we use the parameter {set}
${\bm{\Xi}^{\rm GR}}$ in Eq.~(\ref{eq:theta:GR}). Since the posteriors of
the inclination angle $\iota$ in GWTC-1 have a very broad distribution, we
decide to generate 1000 samples of $\cos\bar\theta_L$ and $\bar\phi_L$
randomly from $[-1, 1]$ and $[0,2\pi)$ respectively. After performing the
Fisher analysis on all of these instances, we list the median value of the
1000 rms errors for each parameter in Table~\ref{tab:PEGR_S}. We do not
include LISA in the table because LISA is not applicable to the Fisher
matrix method due to its low SNRs for stellar-mass BBHs
\citep{Vallisneri:2007ev}.

Except LISA, the angular resolution of the three space-borne detectors span
8 orders of magnitude, from $8.6 \times 10^{-7}\ \rm arcmin^{2}$ in DECIGO
to $66\, \rm arcmin^{2}$ in DO-Conservative. The sky localization by any of
these three space-based detectors is better than the predicted result of
ground GW network observation in \citet{Gaebel:2017zys}. We also found
that, with the full design, the localization of DECIGO is usually $\sim
10^{-5}\ \rm arcmin^{2}$, which is 4--5 orders of magnitude better than the
localization of DOs or a single DECIGO \citep{Nair:2018bxj}, which were
found to be around $1\ \rm arcmin^{2}$. If we had kept $\Phi_0$ the same
for all the constellations of the full DECIGO, the localization would get
worse by 3--4 orders of magnitude. This indicates the importance of the
distribution of the detectors on the ecliptic plane.

We also perform parameter estimation for a 1-year duration. When we compare
the angular resolution of corresponding space-borne detector operating for
1 year with its counterpart for 4 years, we found that DECIGO's result
stays the same, and the results from both DO-Optimal and DO-Conservative
are 2--3 times worse. This indicates on one hand that, the number and
orbital distribution of the detectors make a big difference, and on the
other hand that, decihertz observation does not need too much time to
resolve a source's location, {thus they can be used to warn other detectors
in advance of the coalescence}. {We also found that,} for individual
sources, generally, a source with larger SNR would yield a smaller sky
localization area, but the result depends on the specific location of the
source.

The fractional uncertainty on the distance, $\Delta D_L/D_L$, is about
$10^{-3}$--$10^{-2}$ for DOs and DECIGO. Because the distance information
is encoded in the GW amplitude, the naively expected value of $\Delta
D_L/D_L$ is $\sim 1/\rho$. However, since the 4 directional parameters in
$\bm{\Xi}^{\rm loc}$ are contained in $F^{+,\times}$, thus they also change
the GW amplitude received by the detector. Consequently, when measuring
both distance and direction, the uncertainty in $\Delta D_L/D_L$ will be
larger than the expected $1/\rm \rho$ scaling.

Now we discuss the estimation of the intrinsic parameters for DOs and
DECIGO. The measurement of $\Delta {\cal M}/{\cal M}$ is less than
$10^{-6}$ for all detectors. In DO-Conservative, the value of $\Delta {\cal
M}/{\cal M}$ for different sources ranges from $4.8\times 10^{-9} $
(GW170608) to $1.9 \times 10^{-7}$ (GW170729), which spans about two orders
of magnitude. For a PN inspiral signal, the measurement of the chirp mass
follows the relation \citep{Cutler:1994ys}
\begin{equation}
	\Delta {\cal M}/{\cal M} \propto \rho^{-1} {\cal M}^{5/3}\,.
\end{equation}
This naturally accounts for the two orders of magnitude difference between
GW170608 whose ${\cal M} = 8.5\,M_\odot$ and GW170729 whose ${\cal M} =
53.5\,M_\odot$. The value of $\Delta {\cal M}/{\cal M}$ in DO-Optimal and
DECIGO for each source is approximately 2 and 10 times better than that in
DO-Conservative. In each detector, the precision of $\eta$ is worse by
$\sim$ four orders of magnitude than the precision of ${\cal M}$, but the
values of $\Delta\eta/\eta$ for different sources generally follow the same
trend as the values of $\Delta{\cal M}/{\cal M}$ for them. For the spin
effects, it is $\chi_{\rm PN}$ that dominantly characterizes the IMRPhenomD
waveform, which is defined as
\begin{equation}
	\chi_{\mathrm{PN}}=\chi_{\mathrm{eff}}-\frac{38
	\eta}{113}\left(\chi_{1}+\chi_{2}\right)\,, \label{eq:chipn}
\end{equation}
where
\begin{equation}
	\chi_{\mathrm{eff}} = \frac{m_{1} \chi_{1}+m_{2} \chi_{2}}{M}\,.
\end{equation}
As a result, $\Delta \chi_1$ is smaller than $\Delta \chi_2$ due to the
fact that $m_1 \geq m_2$ and the error propagation formula exerted on
Eq.~(\ref{eq:chipn}). Therefore we treat $\Delta \chi_1$ as a
representative of spin parameter measurement precision and investigate the
variation of it in different observation scenarios. The prior information
makes the dispersion of spin errors relatively small, with $\Delta \chi_1 =
0.062\mbox{--}0.26$ in DO-Conservative, 0.015--0.11 in DO-Optimal, and
0.0018--0.0094 in DECIGO.

\subsection{Parameter estimation with ground-based detectors}

For ground-based detectors, the parameter set we considered is
${\bm{\Xi}^{\rm GR}_{\rm w/o\,loc}}$ in Eq.~(\ref{eq:Xi:noloc}). Since the
duration of the BBH merger in hectohertz band lasts about minutes, we
ignore the detectors' motion with the Earth. The results are shown in
Table~\ref{tab:PEGR_G}. Among all ground-based detectors, CE provides the
most precise measurements, except for the measurement of chirp mass, which
ET measures slightly better because of its lower-frequency sensitivity.
Comparing to the space-borne detectors, the uncertainty on the chirp mass
and the symmetric mass ratio from ground-based detectors are worse, by 2--8
orders of magnitude. This is due to the fact that the mass information is
mostly encoded in the GWs' phasing and {space-based} detection contains
much more numbers of GW cycles than that of {ground-based} detection. The
spin measurement in space-borne detectors is also generally better than
that of ground-based detectors.

On the contrary, without the inclusion of the four directional parameters
in $\bm{\Xi}^{\rm loc}$, the measurement of luminosity distance from the
third generation detectors, CE and ET, is generally better than that of the
space-based detectors. For example, CE's precision on the distance is $\sim
4$ times better than that of DECIGO. Note that the result from ground-based
detection satisfies $\Delta D_L/D_L \sim 1/\rho$ . As for the measurement
of $t_c$, the resolution ability cannot simply be judged in terms of the
detector type being space-based or ground-based. From the lowest precision
to the highest precision of $t_c$ measurement, the order follows from
DO-Conservative ($\Delta t_c \simeq 35\,\rm ms$), DO-Optimal ($\Delta t_c
\simeq 13\,\rm ms$), AdvLIGO ($\Delta t_c \simeq 1.6\,\rm ms$), A+LIGO
($\Delta t_c \simeq 1.4\,\rm ms$), ET ($\Delta t_c \simeq 0.37\,\rm ms$),
CE ($\Delta t_c \simeq 0.23\,\rm ms$), and DECIGO ($\Delta t_c \simeq
0.082\,\rm ms$), where the time precision in the parentheses is the
smallest $t_c$ error measured in each detector among all LIGO/Virgo BBH
sources.

\subsection{Parameter estimation with joint observations}

\begin{table*}
	\centering
	\caption{Parameter estimation in GR for LIGO/Virgo BBHs by the joint
	observation of CE and a space-borne detector. Notice that, due to a low
	SNR in LISA (see Table~\ref{tab:snr}), the location resolution $\Delta
	\Omega$ with LISA is only indicative, and a proper calculation,
	probably via Monte Carlo sampling, is needed for LISA to account for the
	shortcomings of the Fisher matrix \citep{Vallisneri:2007ev,
	Toubiana:2020vtf}.}
	\label{tab:PEGR_joint}
	\begin{tabular}{lllllllllll} 
		\hline\hline
		GW & $\Delta D_L/D_L$ & $\Delta {\cal M}/{\cal M}$ & $\Delta
		\eta/\eta$ & $\Delta t_c$ & $\Delta \phi_c$ & $\Delta \chi_1$ &
		$\Delta \chi_2$ & $\Delta \Omega$ \\
		&($\%$)&$(10^{-8})$& ($10^{-4}$) &(ms)&&&&(${\rm arcmin}^2$)
		 \\
		 \hline
		 LISA + CE \\
GW150914 &  0.042 & 33 & 14 & 0.21 & 0.011 & 0.053 & 0.062 & 1500 \\
GW151012 &  0.15 & 150 & 71 & 0.53 & 0.026 & 0.052 & 0.088 & 110000 \\
GW151226 &  0.10 & 96 & 47 & 0.21 & 0.017 & 0.036 & 0.064 & 44000 \\
GW170104 &  0.11 & 89 & 39 & 0.45 & 0.025 & 0.041 & 0.065 & 22000 \\
GW170608 &  0.080 & 63 & 32 & 0.12 & 0.014 & 0.042 & 0.061 & 160000 \\
GW170729 &  0.19 & 89 & 44 & 1.3 & 0.045 & 0.066 & 0.10 & 25000 \\
GW170809 &  0.099 & 74 & 33 & 0.49 & 0.025 & 0.043 & 0.066 & 15000 \\
GW170814 &  0.062 & 51 & 22 & 0.26 & 0.015 & 0.059 & 0.073 & 4500 \\
GW170818 &  0.098 & 71 & 32 & 0.51 & 0.026 & 0.062 & 0.085 & 15000 \\
GW170823 &  0.15 & 89 & 44 & 0.93 & 0.039 & 0.086 & 0.12 & 64000 \\
\hline
DO-Conservative + CE \\
GW150914 &  0.039 & 2.1 & 1.4 & 0.055 & 0.0034 & 0.014 & 0.016 & 0.050 \\
GW151012 &  0.14 & 3.0 & 2.5 & 0.092 & 0.010 & 0.0079 & 0.015 & 2.1 \\
GW151226 &  0.098 & 0.67 & 0.71 & 0.027 & 0.0064 & 0.0037 & 0.0075 & 0.33 \\
GW170104 &  0.10 & 2.8 & 2.1 & 0.087 & 0.0066 & 0.0073 & 0.012 & 0.46 \\
GW170608 &  0.079 & 0.36 & 0.47 & 0.017 & 0.0048 & 0.0049 & 0.0076 & 0.83 \\
GW170729 &  0.17 & 9.0 & 5.6 & 0.41 & 0.012 & 0.019 & 0.030 & 1.4 \\
GW170809 &  0.092 & 2.7 & 1.9 & 0.087 & 0.0054 & 0.0071 & 0.011 & 0.25 \\
GW170814 &  0.058 & 1.9 & 1.4 & 0.052 & 0.0038 & 0.011 & 0.014 & 0.068 \\
GW170818 &  0.090 & 4.1 & 3.0 & 0.12 & 0.0073 & 0.014 & 0.020 & 0.49 \\
GW170823 &  0.14 & 8.1 & 6.2 & 0.30 & 0.014 & 0.027 & 0.038 & 4.3 \\
\hline
DO-Optimal + CE \\
GW150914 &  0.039 & 0.76 & 0.43 & 0.024 & 0.0014 & 0.0056 & 0.0067 & 0.0059 \\
GW151012 &  0.14 & 1.3 & 0.93 & 0.047   & 0.0053 & 0.0038 & 0.0072 & 0.23 \\
GW151226 &  0.098 & 0.23 & 0.26 & 0.015 & 0.0033 & 0.0019 & 0.0038 & 0.026 \\
GW170104 &  0.10 & 0.94 & 0.60 & 0.045    & 0.0034 & 0.0035 & 0.0059 & 0.049 \\
GW170608 &  0.079 & 0.10 & 0.15 & 0.0098 & 0.0025 & 0.0024 & 0.0038 & 0.058 \\
GW170729 &  0.17 & 4.1 & 2.3 & 0.29   & 0.0050 & 0.013 & 0.021 & 0.17 \\
GW170809 &  0.092 & 0.83 & 0.48 & 0.050 & 0.0030 & 0.0039 & 0.0062 & 0.025 \\
GW170814 &  0.058 & 0.56 & 0.34 & 0.026 & 0.0019 & 0.0052 & 0.0065 & 0.0071 \\
GW170818 &  0.090 & 1.3 & 0.84 & 0.058   & 0.0032 & 0.0064 & 0.0090 & 0.055 \\
GW170823 &  0.14 & 4.1 & 2.4 & 0.15   & 0.0058 & 0.013 & 0.018 & 0.56 \\
\hline
DECIGO + CE \\
GW150914 &  0.033 & 0.12 & 0.067 & 0.011 & 0.00074 & 0.0026 & 0.0031 & 8.5 $\times 10^{-7}$ \\
GW151012 &  0.13 & 0.14 & 0.11 & 0.026 & 0.0031 & 0.0020 & 0.0038 & 4.9 $\times 10^{-5}$ \\
GW151226 &  0.092 & 0.032 & 0.034 & 0.011 & 0.0021 & 0.0012 & 0.0023 & 1.4 $\times 10^{-5}$ \\
GW170104 &  0.089 & 0.18 & 0.12 & 0.024 & 0.0020 & 0.0019 & 0.0031 & 1.0 $\times 10^{-5}$ \\
GW170608 &  0.074 & 0.018 & 0.021 & 0.0059 & 0.0015 & 0.0013 & 0.0020 & 4.5 $\times 10^{-5}$ \\
GW170729 &  0.14 & 0.80 & 0.52 & 0.11 & 0.0027 & 0.0051 & 0.0081 & 1.8 $\times 10^{-5}$ \\
GW170809 &  0.081 & 0.22 & 0.15 & 0.031 & 0.0019 & 0.0024 & 0.0038 & 1.0 $\times 10^{-5}$ \\
GW170814 &  0.051 & 0.13 & 0.084 & 0.015 & 0.0012 & 0.0029 & 0.0037 & 2.6 $\times 10^{-6}$ \\
GW170818 &  0.077 & 0.24 & 0.15 & 0.028 & 0.0017 & 0.0031 & 0.0043 & 7.1 $\times 10^{-6}$ \\
GW170823 &  0.12 & 0.43 & 0.29 & 0.059 & 0.0025 & 0.0049 & 0.0071 & 4.3 $\times 10^{-5}$ \\
		\hline
	\end{tabular}
\end{table*}

\citet{Nair:2018bxj} have analyzed the parameter estimation ability of the
single-detector DECIGO in its geocentric orbit and heliocentric orbit, as
well as its synergy with ET. We here consider the joint observation of all
the space-based detectors in their full configuration with CE.

To estimate parameter precision from joint observation, we add the Fisher
matrix via \citep{Cutler:1994ys}
\begin{equation}
	\Gamma_{ab}^{\mathrm{joint}}
	=\Gamma_{ab}^{\mathrm{space}}+\Gamma_{ab}^{\mathrm{CE}}\,.
\end{equation}
As discussed before, the Fisher matrix of CE which uses $\bm{\Xi}^{\rm GR}_{\rm
w/o\,loc}$ has 7 dimensions, while the Fisher matrix of space-based
detectors which uses $\bm{\Xi}^{\rm GR}$ has 11 dimensions. To combine them, we
augment the ground-based Fisher matrix with elements at the position where
the four directional parameters in $\bm{\Xi}^{\rm loc}$ locate and set
these matrix elements to zero. Therefore in joint observation, we use
parameter set ${\bm{\Xi}^{\rm GR}}$ in Eq.~(\ref{eq:theta:GR}). In this way
the sky localization is still determined by the space-based detectors, but
the adding of the ground-based observation helps to narrow down the errors
of other parameters. For space-borne detectors, we also generate 1000
samples of $\cos\bar\theta_L$ and $\bar\phi_L$ randomly from $[-1, 1]$ and
$[0,2\pi)$ for each BBH. Then we carry out Fisher analysis in joint
observation for all of them and list the median value of the 1000 rms
errors for each parameter in Table~\ref{tab:PEGR_joint}.

Comparing the results in Table~\ref{tab:PEGR_joint} to the results of the
space-borne detection alone in Table~\ref{tab:PEGR_S}, all the precisions
get improved. We will discuss the results in terms of different parameters.
For the angular resolution, the joint observation of CE and DECIGO yields
the smallest sky area of $8.5 \times 10^{-7}\rm \ arcmin^{2}$ for the
source GW150914. Joint observation in general gives a result 1--220 times
better than the space-based observation alone. It is because that the
space-borne observations can inform ground-based detectors the location of
the sources and then the ground-based detections help to better determine
the precisions of other physical parameters, finally leading to a better
angular resolution. We define the enhancement in the sky area measurement,
$ {\cal F} \equiv \Delta \Omega_{\rm space}/\Delta \Omega_{\rm joint}$,
where $\Delta \Omega_{\rm space}$ is the angular resolution estimated from
the space-borne observation alone and $\Delta \Omega_{\rm joint}$ is the
angular resolution estimated by the joint observation of CE and the
corresponding space-borne detector. The enhancement is most significant in
joint observation of CE and DO-Optimal. The best result comes from
GW170729, which gives ${\cal F} \simeq 220$. The average improvement in
DO-Optimal is ${\cal F} \simeq 90$, while the average improvement in
DO-Conservative is ${\cal F} \simeq 20$. DECIGO has an exceptionally low
enhancement, ${\cal F} \leq 1.1$ for all sources. It means {that} CE cannot
provide much more precise measurement than the full DECIGO alone, which is
already by itself very powerful in localization.

The measurement of $D_L$ is dominated by the observation of CE, with an
improvement of less than 25\% from joint observation. In contrast, the
measurements on ${\cal M}\, \rm and\, \eta$ are dominated by space-based
observations. The joint observation of DO-Optimal with CE gives a result
2--4 times better than the DO-Optimal alone, while for DO-Conservative and
DECIGO, the joint estimation on ${\cal M}$ and $\eta$ is 1--2 times better
than the space-based detection alone. Before joint observation, the
estimated spin errors $\Delta \chi_1$ of CE and DO-Optimal have the same
order of magnitude in the range 0.01--0.1. Joint observation of CE and
DO-Optimal strengthens the result by an order of magnitude. Other joint
observations would not have such a great improvement for the spin
measurement. As usual, the joint observation of CE and DECIGO gives the
best limit, $\Delta \chi_1= $ 0.001--0.005.

Furthermore, we also consider the case in which the events are jointly
observed by three detectors across the whole frequency band, i.e. selecting
one detector from one waveband. We found that, as we add the millihertz
``LISA'' to the existing combinations ``DO + CE'' and ``DECIGO + CE'', the
parameter precision only improves at a relative level $\lesssim 10^{-3}$.
It indicates that, for stellar-mass BBHs, when decihertz detectors are
involved, the effect from the addition of millihertz detectors is very
marginal.

Evaluating the improvement of the joint observation on the parameter
estimation as a whole, the combination of CE and DO-Optimal gives the most
prominent enhancement. Therefore the synergy of CE and DO-Optimal is the
best. On the other hand, the full DECIGO is already by itself very powerful
in the parameter estimation. Nonetheless, the combination of CE and DECIGO
always gives the best parameter precision.

\begin{table*}
	\centering
	\caption{Constraints on the dipole radiation parameter $B$ from
	different detectors and their combinations. Observation duration is set
	to 4 years for space-borne detectors. }
	\label{tab:Dipole}
	\begin{tabular}{|l|ll|lll|lll|} 
		\hline
		\hline
		   &AdvLIGO  & CE & LISA &  DO-OPT & DECIGO 
		   & CE\,\&\,LISA & CE\,\&\,DO-OPT & CE\,\&\,DECIGO \\
		   & ($10^{-3}$)  & ($10^{-5}$) & ($10^{-8}$) &  ($10^{-10}$) & ($10^{-12}$) 
		   &  ($10^{-10}$) & ($10^{-12}$) & ($10^{-12}$)\\
		 \hline
GW150914 & 1.3 & 2.0 & 0.59   & 0.60  & 9.4 & 2.4 & 7.9 & 1.3  \\
GW151012 & 1.4 & 1.6 & 2.3    & 0.66 & 7.9  & 7.5  & 13 & 1.9  \\
GW151226 & 0.19 & 0.45 & 2.0  & 0.18 & 1.9 & 5.2& 3.4 &  0.53  \\
GW170104 & 2.3 & 2.9 & 1.3    & 0.77 & 12  & 5.7& 13  &  2.3  \\
GW170608 & 0.11 & 0.28 & 1.3  & 0.099 & 1.1 & 4.4& 2.1 &  0.32  \\
GW170729 & 15 & 36  & 2.6     & 6.9 & 120    & 11  & 69 &  13  \\
GW170809 & 3.0 & 4.4 & 1.1    & 1.0 & 16   & 4.6 & 14 &  2.7  \\
GW170814 & 1.4 & 1.9 & 0.74   & 0.56 & 9.0 & 3.1 & 8.5 & 1.5  \\
GW170818 & 3.3 & 5.3 & 1.2    & 1.3 & 19   & 4.9  & 17 & 2.9  \\
GW170823 & 7.8 & 17 & 2.2     & 3.6 & 46   & 7.6  & 46 & 6.1  \\
		\hline
	\end{tabular}
\end{table*}

\section{Tests of gravity: dipole radiation}
\label{sec:dipole}

\begin{figure}
	\includegraphics[width=8.5cm]{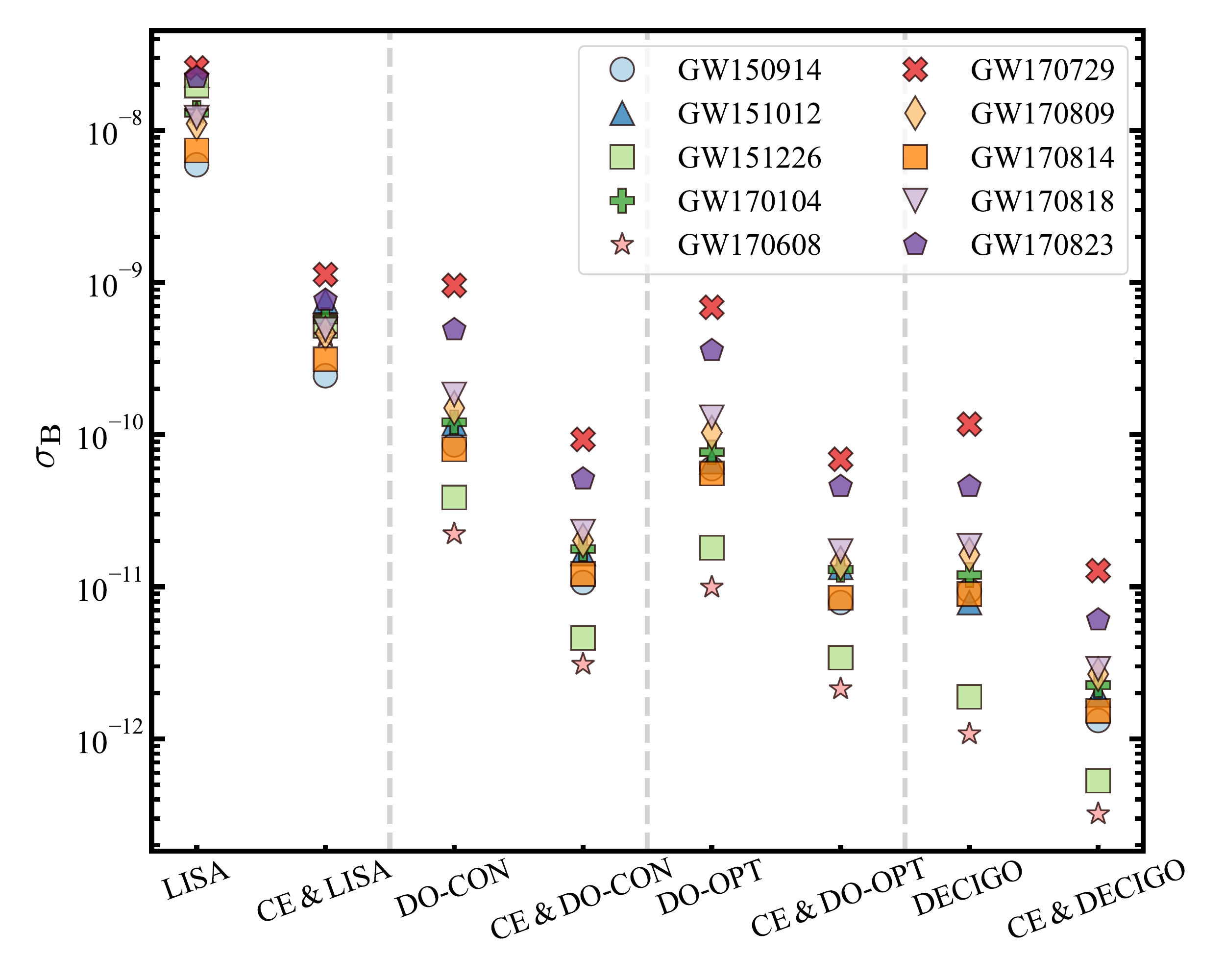}
	\caption{Constraints on the dipole radiation parameter $B$ for all the
	BBH sources in GWTC-1 as a function of different observation scenarios.
	The duration of observation is assumed to be 4 years.}
    \label{fig:dipole}
\end{figure}

Now we present an important example in testing GR with multiband
observation, namely the test of gravitational dipole radiation. GR has been
confirmed by extensive experimental tests and most of these tests are done
in the weak field regime \citep{Will:2014kxa}. However, there are theories
where gravity is significantly modified when the gravitational field is
strong. Such theories predict different structures and dynamical evolutions
of compact objects, as well as binaries that are composed of compact
objects. Pulsar timing and GW observations have put interesting bounds to
them \citep{Berti:2015itd, Shao:2017gwu, Sathyaprakash:2019yqt}. For
example, in a class of scalar-tensor theory, the scalar field is
nonminimally coupled to the Ricci scalar in the physical frame, and for NSs
strong-field effects may appear \citep{Damour:1993hw, Damour:1996ke}. These
effects were constrained tightly by limiting the amount of dipole
radiation via pulsar timing \citep{Freire:2012mg, Wex:2014nva,
Shao:2017gwu, Zhao:2019suc}. However, due to the no-hair theorem, vacuum
spacetime normally does not excite the scalar field, thus this class of
theory is irrelevant to BBH events that we are considering. Nevertheless,
recently it was discovered that some kinds of scalar-tensor theories will
have BBH systems emitting the gravitational dipole radiation
\citep[][and references therein]{Berti:2018cxi}. 

In this study, inspired by specific theories, we adopt a generic
parametrized dipole flux correction for BBHs \citep{Barausse:2016eii},
\begin{equation}\label{eq:Edot}
	\dot{E}_{\mathrm{GW}} =
	\dot{E}_{\mathrm{GR}}\bigg[1+B\left(\frac{M}{r}\right)^{-1}\bigg]\,,
\end{equation}
where $\dot{E}_{\mathrm{GR}}$ is the quadrupole flux in GR, $M$ is the
binary total mass and $r$ is the separation between two BHs. Parameter $B$
is a generic parameter quantifying the magnitude of the dipole radiation.
It varies in different theories. For example, in
Jordan-Fierz-Brans-Dicke-like theories for NSs \citep{Brans:1961sx}, $B =
5(\Delta \alpha)^{2} / 96$, where $\Delta \alpha$ is the difference between
the scalar charges of the two bodies. Note that in Eq.~(\ref{eq:Edot}), we
have $({M}/{r})^{-1} \sim {v}^{-2}$ where $v$ is the relative velocity of
two bodies. Therefore, the dipole radiation corresponds to {$-1$}\,PN
correction.

A quantitative test of GR at the level of the waveform could be conducted
by adding phenomenological corrections to the GR waveform. We adopt the
restricted parametrized post-Einsteinian (ppE) formalism
\citep{Yunes:2009ke} in our calculation. At the inspiral stage, the ppE
model adds general phase correction at different PN orders,
\begin{equation}
	{\tilde{h}}_{\mathrm{ppE}}(f)=A_{\mathrm{GR}}(f)
	e^{i\left[\Psi_{\mathrm{GR}}(f)+\beta u^{b}\right]}\,,
	\label{eq:ppE}
\end{equation}
where $u = \pi {\cal M} f \propto v^3$, $A_{\mathrm{GR}}$ and
$\Psi_{\mathrm{GR}}$ are the GW amplitude and phase in GR\@. Parameter $b$
indicates the frequency dependence of the non-GR correction. For dipole
radiation at $-1$\,PN, $b = -7/3$. Coefficient $\beta$ is the magnitude of
the non-GR phase correction, and connecting to Eq.~(\ref{eq:Edot}) we have
$\beta=-(3 / 224) \eta^{2 / 5} B$. Note that we have ignored the non-GR
modification to the GW amplitude which is subdominant in our analysis
\citep[see e.g.][]{Tahura:2019dgr}.

The ppE formalism has been extensively used to explore various deviations
from GR \citep{Yunes:2016jcc, Chamberlain:2017fjl}. The multiband
observation can further improve the bounds on alternative gravity theories
\citep{Vitale:2016rfr}. Using the Fisher matrix analysis,
\citet{Barausse:2016eii} explored the prospects in constraining $B$ in
dipole radiation from various sources to be observed by AdvLIGO and
different LISA configurations. Recently, \citet{Toubiana:2020vtf} further
performed a full Bayesian analysis to constrain the ppE parameters of BBH
systems to be observed by LISA with or without the prior on the coalescence
time $t_c$ provided by the ground-based observation.
\citet{Gnocchi:2019jzp} and \citet{Carson:2019rda, Carson:2019kkh}
predicted bounds on different alternative theories in multiband
observation, however, the parameter space they explored is exclusive of the
directional parameters in $\bm{\Xi}^{\rm loc}$. Here we take into account
the directional information and investigate the bounds on the generic
{gravitational dipole radiation} parameter $B$.

Including $B$, the parameter set we consider in the ppE model becomes
larger by one dimension. Similar to Section~\ref{sec:pe}, we denote two
sets of parameters used in the following discussion by
\begin{align}
	{\bm{\Xi}^{\rm dipole}_{~~~~~~~~~~}} &= \bm{\Xi}^{\rm GR} \cup \big\{ B
	\big\}\,, \label{eq:Xi:dipole} \\
    {\bm{\Xi}^{\rm dipole}_{\rm w/o\,loc}} &= \bm{\Xi}^{\rm GR}_{\rm
	w/o\,loc} \cup \big\{ B\big\}\,. \label{eq:Xi:dipole:noloc}
\end{align}
As before, we generate 1000 samples with $\cos\bar\theta_L$ and
$\bar\phi_L$ randomly from $[-1, 1]$ and $[0,2\pi)$. We use
Eq.~(\ref{eq:ppE}) and carry out the Fisher matrix analysis on these
samples. We record the median value from the analysis of the 1000 rms
errors of $B$ (denoted as $\sigma_B$) for each BBH system.

\begin{figure*}
	\includegraphics[width=16cm]{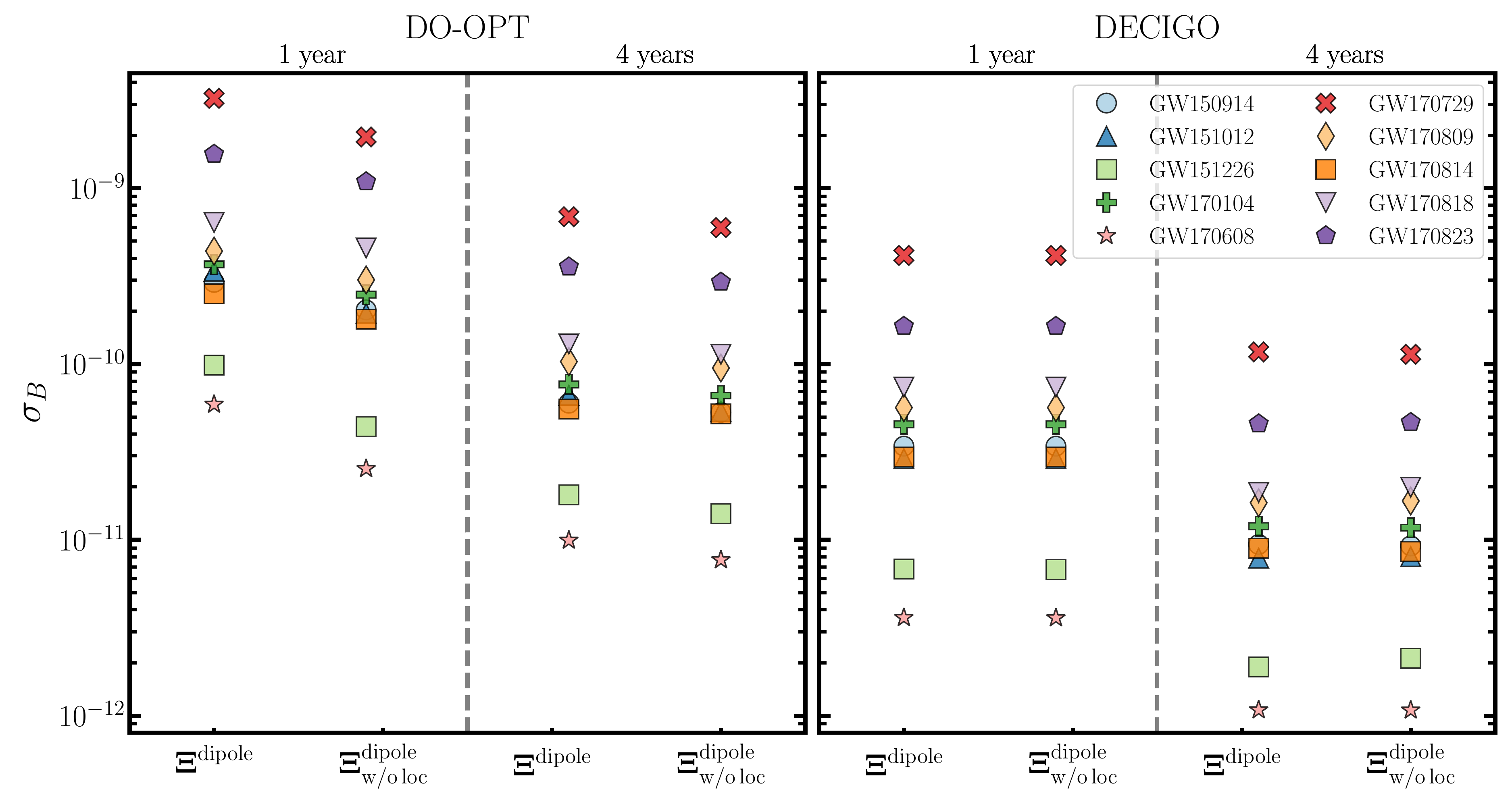}
	\caption{Constraints on the dipole radiation parameter $B$ for all the
	BBH sources in GWTC-1 for {different parameter space sets in the Fisher
	matrix analysis}. Two space-borne detectors, DO-Optimal and DECIGO, are
	considered. For each detector (panel), the region left\,/\,right to the
	dashed line has an observation duration of 1 year\,/\,4 years. For each
	column, the label in the abscissa denotes the choice of the parameter
	set used in the Fisher matrix.}
    \label{fig:dipole_sky}
\end{figure*}

In Table~\ref{tab:Dipole} we list the constraints on {$B$} that we can
obtain from selected ground-based, space-based, and joint observations. We
found that space-based constraints are 3--7 orders of magnitude better than
ground-based constraints, whereas the best constraint $\sigma_B \sim
10^{-12}$ comes from DECIGO\@. This is caused by the combination of large
SNRs and a large number of cycles in the low frequency. Joint observations
have a factor of 3--24 further improvement and the best constraint,
$\sigma_B \sim 3 \times 10^{-13}$, comes from {the combination of CE and
DECIGO}. In Fig.~\ref{fig:dipole} we compare the constraints on $\sigma_B$
between single-detector observation and joint observation of CE and
space-borne detectors. We found that, with the joining of CE, LISA has the
most significant improvement, up to a factor of $\sim24$ for {GW150914-like
sources}; DECIGO has the smallest improvement, about $\sim 3$ for
{GW170608-like sources}.

Among all sources, GW170608 has the lowest redshifted chirp mass, ${\cal M}
\sim 8.5 \, M_\odot$. We notice that once the SNR is high enough, GW170608
always constrains $\sigma_B$ better than other BBHs, for BBHs with lower
${\cal M}$ have a lower {frequency parameter} $u$, which is an advantage in
constraining the negative PN order parameter $B$. If we focus our attention
on DECIGO and {the joint observation of CE and DECIGO}, we found {that} the
improvement of joint observation is most significant for heavy systems. For
example the improvement is about 9.2 for GW170729-like sources. For light
systems like GW170608, only a factor of 3.4 is possible for the
improvement. We have also performed calculation for three-detector joint
observation and, similarly as before, we conclude that for stellar-mass
BBHs the addition of LISA changes the bounds on parameter $B$ very
marginally.

For {space-based} detection, we also investigate the effects from different
observation duration, say $T_{\rm obs} = 1$\,yr versus $T_{\rm obs} =
4$\,yrs, as well as the effects from different parameter sets, namely
$\bm{\Xi}^{\rm dipole}$ and $\bm{\Xi}^{\rm dipole}_{\rm w/o\,loc}$ in
Eq.~(\ref{eq:Xi:dipole}) and Eq.~(\ref{eq:Xi:dipole:noloc}) respectively.
In total we have four combinations of observational scenarios. In
Fig.~\ref{fig:dipole_sky} we plot the results for two space-borne
detectors, DO-Optimal and DECIGO, respectively in the left and right
panels. In each panel, the four cases are divided into two blocks for
1-year observation and 4-year observation. Notice that in each panel, the
third column is the same as what was plotted in Fig.~\ref{fig:dipole}; the
second and fourth columns are the constraints that exclude the
consideration of directional parameters in $\bm{\Xi}^{\rm loc}$, as was
done by \citet{Barausse:2016eii}, \citet{Gnocchi:2019jzp}, and
\citet{Carson:2019kkh}.

In Fig.~\ref{fig:dipole_sky}, comparing the first column with the third
column of each panel, we found that the 4-year observation improves the
constraints on $B$ by a factor of 4.9 and 3.5 for DO-Optimal and DECIGO
respectively. For each detector, comparing the two columns in the 1-year
observation scenario and the 4-year observation scenario, we found that the
inclusion of sky location worsens the constraints on $B$ in DO-Optimal, but
hardly affects the constraints in DECIGO\@. For DECIGO, because it has 8
detectors, the sky location measurement is already independent of the
parameter $B$ measurement, no matter in the 1-year or 4-year observation
scenarios. For DO-Optimal, in the 1-year observation scenario, the
inclusion of sky location at most worsens the constraints on $\sigma_B$ by
a factor of 2.3, while in the 4-year observation scenario, it worsens the
constraints on $\sigma_B$ by a factor of 1.3. Therefore, longer observation
could break the degeneracy between directional parameters and the dipole
parameter $B$ better. Notice that, with the inclusion of the directional
parameters in $\bm{\Xi}^{\rm loc}$, the precision of $B$ for all the
sources has similar fractional drops in spite of different SNRs and chirp
masses.

\section{Discussion}
\label{sec:disc}
 
After the first discovery of GWs by the LIGO/Virgo Collaboration with
ground-based GW detectors, space-based GW astronomy is guaranteed to
trigger a new leap in the field for the next decade. Among the many
promising targets, the synergy between ground-based and space-based GW
detectors, namely the the multiband observation, will provide unprecedented
science ever~\citep{Sesana:2016ljz,Cutler:2019krq}. In this work, we
investigate the signal characteristics and parameter estimation for GWTC-1
BBHs with future ground-based and space-based detectors. We show that the
multiband observation of CE and space-borne detectors improves the
precision of all the considered parameters from a few percent to two orders
of magnitude. With its full design and for a 4-year observational time,
DECIGO will generate the smallest sky localization area of around $10^{-5}\
\rm arcmin^{2}$ for stellar-mass BBHs. The joint observation of DECIGO and
CE gives the best estimation on the parameter precision, while the joint
observation of DO-Optimal and CE has the most prominent enhancement,
compared to the corresponding space-based observation alone.

The catalog GWTC-1 was used to perform tests of gravity \citep[see
e.g.][]{LIGOScientific:2019fpa, Shao:2020shv}. We here study the projected
constraints on the gravitational dipole radiation in different observation
scenarios. Differently from early study, but similarly to a recent one for
LISA by \citet{Toubiana:2020vtf}, we take into consideration of the orbital
motion for the space-borne detectors. We found that the best constraint on
a generic dimensionless dipole parameter, $\sigma_B \sim 10^{-12}$, comes
from the joint observation of DECIGO and CE\@. Finally we compare the
bounds on the parameter $B$ with different choices of parameter sets and
observational time for DO-Optimal and DECIGO\@. We conclude that the
degeneracy between the directional parameters (including those that
characterize the sky location and orbital orientation of the BBH) and the
dipole radiation parameter $B$ can be broken by increasing the
observational time. This kind of
work can be easily extended to other alternative gravity theories within
the ppE framework, for examples the Einstein-dilaton-Gauss-Bonnet theory
and the dynamical Chern-Simons gravity, as was considered by
\citet{Gnocchi:2019jzp}.

The smaller sky localization area and deeper distance range provided by the
multiband observation can help electromagnetic (EM) telescopes with narrow
field-of-view to better trace the GW sources \citep{Gehrels:2015uga}.
Although stellar-mass BBH systems are generally believed not to have EM
counterparts, the follow-up can be used to study some exotic scenarios
\citep[see e.g.][]{Ford:2019nic, Gong:2019aqa}. Future multiband and
multimessenger observation will enable us to examine the properties of BBHs
and BH spacetime in a better way than single detectors can do.

Since the LIGO/Virgo BBH signals in LISA have relatively low SNRs, using
ground-based confident detections to identify (in particular sub-threshold)
events from space-based observatories is a new avenue for GW detection.
Such a reverse tracing can increase event rates by a factor of 4--7
\citep{Wong:2018uwb}. This could be a crucial increase for stellar-mass BBH
observations with LISA. Furthermore, a more realistic case where the joint
observation consists of a space-borne detector and a ground-based detector
network (e.g.\ a network consisting of A+LIGO, CE and ET) could also be
considered in future studies \citep{Cutler:2019krq}.

Overall, for stellar-mass BBHs, especially the loudest events among them,
multiband observations from millihertz, decihertz to hectohertz will
enhance and complement parameter estimation derived solely by space-based
or ground-based detection, yield stringent tests of GR, and provide in time
the most accurate sky localization. Therefore, multiband detections in
future enable us to examine fundamental physics of BHs and gravity
theories; meanwhile, with the information of BBH's population in different
GW bands and the possible EM followups, we will better understand the
formation and evolution of BBH systems. In a decade time comes the era of
multiband GW astronomy and its promising future will deliver due scientific
payoff.

\section*{Acknowledgements}

We thank the anonymous referee for helpful comments, and the LIGO/Virgo
Collaboration for providing the posterior samples of their
parameter-estimation studies. This work was supported by the National
Natural Science Foundation of China (11975027, 11991053, 11721303), the
Young Elite Scientists Sponsorship Program by the China Association for
Science and Technology (2018QNRC001), and the High-performance Computing
Platform of Peking University. It was partially supported by the Strategic
Priority Research Program of the Chinese Academy of Sciences through the
Grant No. XDB23010200.


\bibliographystyle{mnras}
\bibliography{refs} 



\appendix

\section{Parameter estimation in individual GW detectors}

In Tables~\ref{tab:PEGR_S} and \ref{tab:PEGR_G} we list our parameter
estimation results for space-based and ground-based GW detectors,
respectively.

\begin{table*}
	\centering
	\caption{Parameter estimation in GR for LIGO/Virgo BBHs in different
	space-based GW detectors. We do not list the results for LISA, due to
	its low SNRs and hence the problem to use the Fisher matrix method.}
	\label{tab:PEGR_S}
	\begin{tabular}{lllllllll} 
		\hline\hline
		GW & $\Delta D_L/D_L$ & $\Delta  {\cal M}/{\cal M}$  & $\Delta \eta/\eta$ & $\Delta t_c$ & $\Delta \phi_c$ & $\Delta \chi_1$ & $\Delta \chi_2$ & $\Delta \Omega$ \\
		
		&(\%)&$(10^{-8})$& $(10^{-4})$ &(ms)&&&&(${\rm arcmin}^2$)
		 \\
		 \hline
DO-Conservative \\
GW150914 &  1.6 & 3.7 & 2.2 & 53 & 0.065 & 0.092 & 0.11 & 0.80 \\
GW151012 &  6.1 & 4.9 & 4.2 & 110 & 0.14 & 0.12 & 0.23 & 15 \\
GW151226 &  3.7 & 0.97 & 1.2 & 48 & 0.071 & 0.070 & 0.14 & 1.5 \\
GW170104 &  3.7 & 5.4 & 3.8 & 100 & 0.13 & 0.097 & 0.16 & 6.9 \\
GW170608 &  2.8 & 0.48 & 0.72 & 35 & 0.058 & 0.062 & 0.096 & 4.5 \\
GW170729 &  5.0 & 19 & 13 & 380 & 0.37 & 0.19 & 0.30 & 57 \\
GW170809 &  3.1 & 6.0 & 4.2 & 120 & 0.13 & 0.10 & 0.17 & 7.4 \\
GW170814 &  2.2 & 3.6 & 2.3 & 77 & 0.082 & 0.10 & 0.13 & 1.8 \\
GW170818 &  3.5 & 7.3 & 5.1 & 120 & 0.13 & 0.16 & 0.23 & 8.8 \\
GW170823 &  5.7 & 15 & 11 & 240 & 0.26 & 0.26 & 0.37 & 66 \\
\hline
DO-Optimal \\
GW150914 &  0.82 & 2.0 & 0.95 & 39 & 0.052 & 0.029 & 0.034 & 0.45 \\
GW151012 &  3.9 & 2.6 & 1.9 & 67 & 0.12 & 0.042 & 0.080 & 6.2 \\
GW151226 &  2.0 & 0.42 & 0.49 & 20 & 0.048 & 0.025 & 0.051 & 0.32 \\
GW170104 &  2.0 & 2.9 & 1.6 & 69 & 0.097 & 0.024 & 0.04 & 3.4 \\
GW170608 &  1.4 & 0.22 & 0.25 & 13 & 0.034 & 0.015 & 0.023 & 0.74 \\
GW170729 &  2.7 & 12 & 6.1 & 280 & 0.28 & 0.056 & 0.090 & 38 \\
GW170809 &  1.6 & 3.4 & 1.8 & 89 & 0.10 & 0.027 & 0.043 & 4.1 \\
GW170814 &  1.1 & 2.0 & 0.99 & 57 & 0.066 & 0.025 & 0.031 & 0.93 \\
GW170818 &  1.9 & 4.1 & 2.3 & 92 & 0.11 & 0.048 & 0.067 & 4.9 \\
GW170823 &  4.0 & 8.8 & 5.4 & 190 & 0.23 & 0.11 & 0.15 & 40 \\
\hline
DECIGO \\
GW150914 &  0.096 & 0.23 & 0.11 & 0.082 & 0.0079 & 0.0038 & 0.0045 & 8.6 $\times 10^{-7}$ \\
GW151012 &  0.47 & 0.32 & 0.21 & 0.31 & 0.016 & 0.0038 & 0.0072 & 5.4 $\times 10^{-5}$ \\
GW151226 &  0.30 & 0.060 & 0.057 & 0.21 & 0.0070 & 0.0026 & 0.0052 & 1.5 $\times 10^{-5}$ \\
GW170104 &  0.29 & 0.39 & 0.22 & 0.20 & 0.016 & 0.0033 & 0.0055 & 1.1 $\times 10^{-5}$ \\
GW170608 &  0.22 & 0.033 & 0.031 & 0.12 & 0.0050 & 0.0018 & 0.0028 & 5.0 $\times 10^{-5}$ \\
GW170729 &  0.38 & 1.5 & 0.88 & 0.54 & 0.047 & 0.0094 & 0.015 & 1.8 $\times 10^{-5}$ \\
GW170809 &  0.27 & 0.50 & 0.30 & 0.26 & 0.016 & 0.0052 & 0.0083 & 1.2 $\times 10^{-5}$ \\
GW170814 &  0.16 & 0.28 & 0.15 & 0.13 & 0.010 & 0.0048 & 0.0060 & 2.9 $\times 10^{-6}$ \\
GW170818 &  0.23 & 0.47 & 0.27 & 0.20 & 0.017 & 0.0056 & 0.0079 & 7.4 $\times 10^{-6}$ \\
GW170823 &  0.32 & 0.84 & 0.49 & 0.35 & 0.029 & 0.0085 & 0.012 & 4.7 $\times 10^{-5}$ \\
		\hline
	\end{tabular}
\end{table*}

\begin{table*}
	\centering
	\caption{Parameter estimation in GR for LIGO/Virgo BBHs in different ground-based GW detectors.}
	\label{tab:PEGR_G}
	\begin{tabular}{llllllll} 
		\hline\hline
		GW & $\Delta D_L/D_L$ & $\Delta  {\cal M}/{\cal M}$  & $\Delta \eta/\eta$ & $\Delta t_c$ & $\Delta \phi_c$ & $\Delta \chi_1$ & $\Delta \chi_2$  \\
		
		&($\%$)&($\%$)&($\%$)  &(ms)&&&
		 \\
		\hline
		AdvLIGO \\
GW150914 &  2.0 & 0.23 & 2.0 & 2.5 & 1.4 & 0.62 & 0.74  \\
GW151012 &  7.3 & 0.26 & 11 & 5.1 & 4.3 & 0.48 & 0.85  \\
GW151226 &  5.0 & 0.11 & 13 & 2.7 & 3.2 & 0.48 & 0.84  \\
GW170104 &  5.3 & 0.36 & 6.2 & 5.7 & 3.3 & 0.51 & 0.81  \\
GW170608 &  3.9 & 0.084 & 11 & 1.6 & 2.7 & 0.56 & 0.80  \\
GW170729 &  11  & 2.5 & 6.8 & 11 & 6.9 & 0.56 & 0.83  \\
GW170809 &  4.9 & 0.46 & 5.2 & 6.1 & 3.2 & 0.53 & 0.81  \\
GW170814 &  3.0 & 0.24 & 3.1 & 2.8 & 1.9 & 0.61 & 0.76  \\
GW170818 &  4.8 & 0.51 & 4.5 & 4.8 & 3.2 & 0.58 & 0.79  \\
GW170823 &  7.8 & 1.1 & 5.9 & 6.4 & 5.3 & 0.59 & 0.80  \\
\hline
A+LIGO \\
GW150914 &  1.0 & 0.16 & 1.3 & 2.3 & 1.0 & 0.58 & 0.68  \\
GW151012 &  3.6 & 0.17 & 6.0 & 4.6 & 2.9 & 0.43 & 0.77  \\
GW151226 &  2.4 & 0.065 & 7.8 & 2.4 & 2.1 & 0.41 & 0.74  \\
GW170104 &  2.6 & 0.24 & 3.8 & 5.1 & 2.4 & 0.46 & 0.73  \\
GW170608 &  1.9 & 0.048 & 6.6 & 1.4 & 1.8 & 0.51 & 0.74  \\
GW170729 &  6.0 & 1.7 & 4.0 & 11 & 5.1 & 0.55 & 0.82  \\
GW170809 &  2.4 & 0.32 & 3.3 & 5.6 & 2.4 & 0.48 & 0.74  \\
GW170814 &  1.5 & 0.16 & 1.9 & 2.6 & 1.4 & 0.57 & 0.70  \\
GW170818 &  2.4 & 0.35 & 2.7 & 4.6 & 2.3 & 0.55 & 0.75  \\
GW170823 &  3.9 & 0.78 & 3.3 & 6.3 & 3.8 & 0.58 & 0.79  \\
\hline
CE \\
GW150914 &  0.043 & 0.0023 & 0.15 & 0.23 & 0.013 & 0.058 & 0.069  \\
GW151012 &  0.16 & 0.0057 & 0.90 & 0.72 & 0.037 & 0.069 & 0.12  \\
GW151226 &  0.11 & 0.0035 & 1.3 & 0.38 & 0.039 & 0.069 & 0.11  \\
GW170104 &  0.11 & 0.0045 & 0.47 & 0.57 & 0.029 & 0.052 & 0.082  \\
GW170608 &  0.089 & 0.0029 & 1.1 & 0.26 & 0.034 & 0.098 & 0.13  \\
GW170729 &  0.19 & 0.028 & 0.54 & 1.7 & 0.11 & 0.086 & 0.13  \\
GW170809 &  0.10 & 0.0046 & 0.38 & 0.55 & 0.030 & 0.049 & 0.074  \\
GW170814 &  0.065 & 0.0029 & 0.26 & 0.33 & 0.017 & 0.073 & 0.090  \\
GW170818 &  0.10 & 0.0050 & 0.35 & 0.55 & 0.031 & 0.067 & 0.091  \\
GW170823 &  0.16 & 0.011 & 0.48 & 0.98 & 0.060 & 0.091 & 0.13  \\
\hline
ET-D \\
GW150914 &  0.11 & 0.0019 & 0.31 & 0.54 & 0.18 & 0.14 & 0.16  \\
GW151012 &  0.42 & 0.0038 & 1.5 & 1.2 & 0.74 & 0.11 & 0.19  \\
GW151226 &  0.37 & 0.0029 & 1.7 & 0.61 & 0.80 & 0.11 & 0.18  \\
GW170104 &  0.29 & 0.0036 & 0.89 & 1.2 & 0.50 & 0.11 & 0.17  \\
GW170608 &  0.33 & 0.0020 & 1.3 & 0.37 & 0.70 & 0.13 & 0.19  \\
GW170729 &  0.50 & 0.011 & 1.1 & 3.7 & 0.60 & 0.18 & 0.28  \\
GW170809 &  0.27 & 0.0039 & 0.77 & 1.3 & 0.43 & 0.11 & 0.17  \\
GW170814 &  0.17 & 0.0023 & 0.47 & 0.67 & 0.29 & 0.15 & 0.18  \\
GW170818 &  0.26 & 0.0041 & 0.72 & 1.3 & 0.41 & 0.16 & 0.22  \\
GW170823 &  0.41 & 0.0070 & 1.0 & 2.4 & 0.55 & 0.22 & 0.30  \\
\hline
	\end{tabular}
\end{table*}

\bsp	
\label{lastpage}
\end{document}